\documentclass[12pt,reqno]{amsart}
\usepackage[left=1in,right=1in,top=1in,bottom=1in]{geometry}%

\usepackage{amsrefs}
\usepackage{amsmath}
\usepackage{enumerate}
\usepackage{amssymb}                
\usepackage{amsmath}                
\usepackage{amsfonts}
\usepackage{amsthm}
\usepackage{bbm}
\usepackage{float}
\usepackage{mathtools}
\usepackage{graphicx}
\usepackage{hyperref}
\usepackage{paralist}

\RequirePackage{xcolor}[2.11]
\colorlet{siaminlinkcolor}{green!50!black}
\colorlet{siamexlinkcolor}{red!50!black}
\colorlet{siamreviewcolor}{black!50}
\hypersetup{
  colorlinks = false,
  allcolors = siaminlinkcolor,
  urlcolor = siamexlinkcolor,
}

\usepackage[capitalize,nameinlink]{cleveref}
\crefname{section}{section}{sections}
\crefname{subsection}{subsection}{subsections}
\Crefname{section}{Section}{Sections}
\Crefname{subsection}{Subsection}{Subsections}

\Crefname{figure}{Figure}{Figures}

\crefformat{equation}{\textup{#2(#1)#3}}
\crefrangeformat{equation}{\textup{#3(#1)#4--#5(#2)#6}}
\crefmultiformat{equation}{\textup{#2(#1)#3}}{ and \textup{#2(#1)#3}}
{, \textup{#2(#1)#3}}{, and \textup{#2(#1)#3}}
\crefrangemultiformat{equation}{\textup{#3(#1)#4--#5(#2)#6}}%
{ and \textup{#3(#1)#4--#5(#2)#6}}{, \textup{#3(#1)#4--#5(#2)#6}}{, and \textup{#3(#1)#4--#5(#2)#6}}

\Crefformat{equation}{#2Equation~\textup{(#1)}#3}
\Crefrangeformat{equation}{Equations~\textup{#3(#1)#4--#5(#2)#6}}
\Crefmultiformat{equation}{Equations~\textup{#2(#1)#3}}{ and \textup{#2(#1)#3}}
{, \textup{#2(#1)#3}}{, and \textup{#2(#1)#3}}
\Crefrangemultiformat{equation}{Equations~\textup{#3(#1)#4--#5(#2)#6}}%
{ and \textup{#3(#1)#4--#5(#2)#6}}{, \textup{#3(#1)#4--#5(#2)#6}}{, and \textup{#3(#1)#4--#5(#2)#6}}

\crefdefaultlabelformat{#2\textup{#1}#3}

\def\noi{\noindent}

\def\R{{\mathbb R}}

\def\C{{\mathbb C}}

\def\calH{{\mathcal H}}
\def\calE{{\mathcal E}}
\def\calQ{{\mathcal Q}}
\def\calL{{\mathcal L}}
\def\calI{{\mathcal I}}
\def\calJ{{\mathcal J}}
\def\calF{{\mathcal F}}

\DeclareMathOperator{\spn}{span}

\newtheorem{lemma}{Lemma}
\newtheorem{theorem}{Theorem}
\newtheorem{corollary}{Corollary}

\newtheorem{hypothesis}{Hypothesis}
\newtheorem{remark}{Remark}

\begin{document}

\title[Multi-pulse solitary waves in a fourth-order NLS equation]{Multi-pulse solitary waves in a fourth-order nonlinear {S}chr{\"o}dinger equation}

\author{Ross Parker}
\address{Department of Mathematics, Southern Methodist University, Dallas, Texas 75275}
\email{rhparker@smu.edu}

\author{Alejandro Aceves}
\address{Department of Mathematics, Southern Methodist University, Dallas, Texas 75275}
\email{aaceves@smu.edu}

\begin{abstract}
In the present work, we consider the existence and spectral stability of multi-pulse solitary wave solutions to a nonlinear Schr\"odinger equation with both fourth and second-order dispersion terms. We first give a criterion for the existence of a single solitary wave solution in terms of the coefficients of the dispersion terms, and then show that a discrete family of multi-pulse solutions exists which is characterized by the distances between the individual pulses. We then reduce the spectral stability problem for these multi-pulses to computing the determinant of a matrix which is, to leading order, block diagonal. Under additional assumptions, which can be verified numerically and are sufficient to prove orbital stability of the primary solitary wave, we show that all multi-pulses are spectrally unstable. Numerical computations for the spectra of double pulses are presented which are in good agreement with our analytical results. Results of timestepping simulations are also provided to characterize the nature of these instabilities.
\end{abstract}

\maketitle

\section{Introduction}

It has been nearly 50 years since the discovery by Zakharov and Shabat \cite{Zak72} of the integrability of the nonlinear Schr\"odinger equation (NLS) and the corresponding soliton solutions, and 40 years since the first experimental demonstration by Mollenauer, Stolen and Gordon of optical solitons propagating in fibers \cite{Moll80}. At its most fundamental level, the NLS soliton represents the balance of chromatic second order dispersion and the Kerr self-focusing nonlinearity. The robustness of the soliton opened up new directions in both theoretical and experimental fronts that continue to this day. Novel fiber designs and technological advances, which led in particular to the invention of the photonic crystal fiber (PCF) and waveguides on a chip, have enabled the engineering of the chromatic dispersion resulting in new discoveries and dynamics beyond the NLS regime; most notably is the generation of supercontinuum in a PCF. In this vein, recent experimental work in silicon photonic crystal waveguides has produced for the first time what is now known as pure quartic optical solitons (PQS) on a chip \cite{BlancoPQS}. The term ``quartic'' indicates that for this waveguide, the leading order dispersion is fourth order. Spectral stability of PQS was shown numerically by Tam et al. \cite{Tam2019}, as well as evolution into PQS from Gaussian initial conditions. This is extended to a more general model in \cite{Tam2020} which contains both second and fourth order dispersion terms, as it is difficult to obtain either pure quartic or pure quadratic dispersion experimentally. Our results presented here provide a more rigorous study of this general model, including, for the first time, results on the existence and spectral stability of multi-pulse solutions.

Multi-pulses, which are multi-modal solitary waves resembling multiple, well-separated copies of a single solitary wave, have been an object of mathematical interest for many years. One notable early example is the demonstration of stationary multi-solitons in a generalized KdV equation by Gorshkov et al. \cite{Gorshkov1979}. The asymptotic method for constructing these multi-pulses was further developed in \cite{Gorskkov1981}, their stability is discussed in \cite{Buryak1997}, and the specific case of existence and stability of double pulses in a 5th order KdV equation is studied in \cite{Pelinovsky2007}. Existence of multi-pulse solutions to a family of fourth-order, reversible Hamiltonian equations was shown in \cite{Buffoni1996} using the dynamics on the Smale horseshoe set, and a spatial dynamics approach to the same problem is found in \cite{SandstedeStrut}.

Of particular interest is the stability of these multi-pulse structures in the case where the primary solitary wave is orbitally stable. A first step is to study the spectrum of the linearization of the the underlying PDE about these solutions. As a motivating example, if the underlying PDE admits a single continuous symmetry, such as translation or gauge invariance, the spectrum of the primary pulse will contain an eigenvalue at the origin. The spectrum of an $n$-pulse will then contain a finite set of eigenvalues near the origin \cite{Alexander1990,Sandstede1998}, one of which will remain at the origin due to the symmetry. Since these additional eigenvalues result from nonlinear interactions between the tails of neighboring pulses, we call them interaction eigenvalues. If the system is non-Hamiltonian, e.g. a reaction-diffusion equation, an $n$-pulse will generally have $n-1$ interaction eigenvalues \cite{Sandstede1998}, whereas if the system is Hamiltonian, there will typically be $n-1$ pairs of interaction eigenvalues \cite{Kapitula2020, Parker2020,Pelinovsky2007} due to the additional eigenvalue symmetry \cite{Kapitula2013} (specific examples are considered below). Some of these interaction eigenvalues may move into the right half plane and cause the multi-pulse to become unstable. As long as the essential spectrum does not enter the right half-plane, spectral stability of multi-pulses is determined by these interaction eigenvalues.

For semilinear parabolic equations with a single eigenvalue at 0 from translation symmetry, these eigenvalues are computed in \cite{Sandstede1998} using Lin's method \cite{Lin1990}, an implementation of the Lyapunov-Schmidt technique, to reduce the problem of finding the eigenvalues near the origin associated with an $n$-pulse to finding nontrivial solutions of an $n\times n$ matrix equation. To leading order, this is equivalent to finding the eigenvalues of an $n \times n$ matrix. Solving this matrix eigenvalue problem yields $n$ eigenvalues near the origin, one of which remains at 0 due to translation invariance. This method is extended in \cite{Manukian} to systems with two continuous symmetries, translation and phase invariance, for which the cubic-quintic complex Ginzburg–Landau equation is a prototypical example. For this system, Lin's method reduces the eigenvalue problem to a $2n\times 2n$ matrix equation. It follows that an $n$ pulse has $2n$ eigenvalues near the origin, two of which remain at 0 from the two symmetries. In certain Hamiltonian systems with a single continuous symmetry, such as the discrete nonlinear Schr{\"o}dinger equation \cite{Parker2020} and a fourth-order beam equation \cite{Kapitula2020}, it has been shown that each additional pulse in a multi-pulse structure gives rise to a pair of interaction eigenvalues. In these cases, an $n$-pulse has $2n$ eigenvalues near the origin; in addition to an eigenvalue at 0 with algebraic multiplicity 2 from the symmetry, there are $n-1$ pairs of nearby interaction eigenvalues which are either real or purely imaginary. A similar result has been shown for double pulse solutions in the fifth-order KdV equation \cite{Pelinovsky2007}. The fourth order nonlinear Schr{\"o}dinger model we consider in this paper is Hamiltonian and has the same two continuous symmetries as in \cite{Manukian}. The analysis combines the approaches in \cite{Manukian} and \cite{Parker2020} to show that an $n$-pulse has $4n$ eigenvalues near the origin; in addition to an eigenvalue at 0 with algebraic multiplicity 4 from the two symmetries, there are $2(n-1)$ pairs of interaction eigenvalues near the origin, which are either real or purely imaginary. Under additional assumptions, which are also used to prove orbital stability of the primary solitary wave, we show that there are $n-1$ pairs of real interaction eigenvalues and $n-1$ pairs of imaginary interaction eigenvalues. Since all multi-pulses have an eigenvalue with positive real part, they are all unstable.

This paper is organized as follows. After a background section on the equation of interest, we present results on the existence and stability of the primary soliton solution in the general case where both second and fourth order dispersion are included. This includes the particular case of PQS. The second part of the paper concerns multi-pulses. We first prove their existence, and then study their spectral stability. We then present numerical examples, both spectral computations and timestepping simulations, and this is followed by a brief discussion of conclusions and directions for future work. The final section contains the proofs for the spectral stability results.

\section{Background}

The following fourth-order generalization of the nonlinear Schr{\"o}dinger equation (NLS) with cubic nonlinearity
\begin{equation}\label{NLS4}
i u_t + \frac{\beta_4}{24}u_{xxxx} - \frac{\beta_2}{2}u_{xx} + \gamma |u|^2 u = 0
\end{equation}
was recently investigated in \cite{Tam2020} in a study of the properties of solitary wave solutions under a combination of second and fourth-order dispersion. (We use the independent variables $(t, x)$ in place of $(z, \tau)$, which is used in \cite{BlancoPQS,Tam2019,Tam2020} and is common in the optics literature). Classical NLS solitons are solutions when $\beta_2 < 0$ and $\beta_4 = 0$. Pure quartic solitons (PQS) occur when $\beta_2 = 0$ and $\beta_4 < 0$, in which case $u(x,t)$ satisfies the equation
\begin{equation}\label{PQSeq}
i u_t + \frac{\beta_4}{24}u_{xxxx} + \gamma |u|^2 u = 0.
\end{equation}
Unlike ordinary NLS solitons, PQS have oscillatory, exponentially decaying tails. Fourth-order nonlinear Schr{\"o}dinger equations have been an object of interest for many years. Karpman and Shagalov \cite{Karpman1996,Karpman1997,Karpman2000} introduced the equation $i u_t + \epsilon u_{xxxx} + u_{xx} + |u|^{2 \sigma}u = 0$ to account for the role of small fourth-order dispersion terms in the propagation of intense laser beams in a bulk medium with Kerr nonlinearity. It has standing wave solutions (known as waveguide solutions in the nonlinear optics literature) which are stable when $\sigma \leq 2$ and $\epsilon < 0$ and unstable when $\sigma \geq 4$. Further results concerning the existence of standing wave solutions are presented in \cite{Bonheure2014,Bonheure2018}, and some global existence results can be found in \cite{Ilan-2002,Shangbi-2007,Pausader2009}. Exact solutions for specific parameters are given in \cite{AbdulMaji2006,KarllsonHook}, and orbital stability of one of these solutions is proved in \cite{Natali2015}. There has been much recent interest in PQS due to the their discovery in experimental media by Blanco-Redondo et al. in 2016 \cite{BlancoPQS}. The existence and spectral stability of PQS solutions was shown numerically in \cite{Tam2019}, and the existence of solitary wave solutions to the more general equation \cref{NLS4} in terms of the parameters $\beta_2$ and $\beta_4$ is discussed in \cite{Tam2020}. We consider only the case $\beta_4 < 0$, which is the regime for which PQS exist \cite{Tam2019} and which is considered in \cite{Tam2020}.

Real-valued, standing wave solutions, i.e. solutions of the form $e^{i \omega t} u(x)$ with $\omega > 0$, satisfy the ODE 
\begin{equation}\label{standingwavereal}
\frac{\beta_4}{24}u_{xxxx} - \frac{\beta_2}{2}u_{xx} + \gamma u^3 - \omega u = 0,
\end{equation}
which is a rescaling of \cite{champneys1998}. For PQS, equation \cref{standingwavereal} can be written in parameter free form by using the rescaling $
u(x; \omega) = \sqrt{\frac{\omega}{\gamma}} \tilde{u}
\left( \left(\frac{\omega}{|\beta_4|}\right)^{1/4}x \right)
$
to obtain the equation
\begin{equation}\label{PQSparfree}
-\frac{1}{24}\tilde{u}_{xxxx} + \tilde{u}^3 - \tilde{u} = 0.
\end{equation} 
We observe that the power or photon number of PQS scales as $\omega^{3/4}$ compared to the $\omega^{1/2}$ scaling of classical NLS solitons. For $\beta_4 < 0$, the more general rescaling \cite[Section VI]{Tam2020} 
\begin{equation}\label{rescaling}
U = \sqrt{\frac{\gamma}{\omega}} U, \quad
X = \left( \frac{24 \omega}{|\beta_4|}\right)^{1/4} x
\end{equation}
transforms equation \cref{standingwavereal} into the one-parameter equation
\begin{equation}\label{eq:NLS4rescaled}
U'''' + 2 \sigma U" +  U - U^3 = 0,
\end{equation}
where $U'$ denotes differentiation with respect to $X$, and 
\begin{equation}\label{eq:sigma}
\sigma = \sqrt{\frac{3}{2 \omega |\beta_4| }}\beta_2
\end{equation}
is a non-dimensional parameter characterizing the relative strengths of the quadratic and quartic dispersion terms. In particular, $\sigma = 0$ for PQS.

For ordinary NLS, an analytic solution can be obtained by the inverse scattering transform \cite{Zak72}. For $\beta_4 < 0$ and $\beta_2 < 0$, an analytic solution has been obtained by Karlsson and H{\"o}{\"o}k \cite{KarllsonHook} when $\omega = 24 \beta_2^2 / 25 |\beta_4|$, which corresponds to $\sigma = -5/4$ in \cref{eq:NLS4rescaled}. The stability of the Karlsson-H{\"o}{\"o}k solution for $\beta_4 = -24$, $\beta_2 = -2$, and $\omega = 4/25$ is discussed in \cite{Natali2015}.

\begin{remark}
In a recent conference presentation \cite{Runge2020}, it was shown that by incorporating an intracavity programmable pulse-shaper in a mode-locked fiber laser, one can manipulate the net cavity dispersion by applying a phase to the pulse so that, to leading order, the stationary pulse generated is modeled by the higher-order NLS equation
\begin{equation}\label{HONLS}
-(i)^k\frac{d^k u}{dx^k}+ \omega u - \gamma u^3 = 0,
\end{equation}
where $k \geq 6$ is a positive, even integer, and the pulse profile satisfies the scaling relation $u(x; \omega) = \sqrt{\frac{\omega}{\gamma}}\tilde{u}(\omega^{1/k}x)$.
\end{remark}

\section{Mathematical setup}\label{sec:setup}

Our analysis follows \cite{Grillakis1990} and \cite[Section 5.2]{Kapitula2013}. Separating real and imaginary parts, equation \cref{NLS4} can be written in standard Hamiltonian form as 
\begin{equation}\label{NLSHam}
\frac{\partial u}{\partial t} = J \calE'(u(t)),
\end{equation}
where $u = (v, w)^T$, $J$ is the symplectic matrix
\begin{equation}
J = \begin{pmatrix}
0 & 1 \\ -1 & 0
\end{pmatrix}
\end{equation}
with $J^2 = -I$, and the energy $\calE$ is given by
\begin{equation}\label{defH}
\calE(u) = \frac{1}{2} \int_{-\infty}^\infty \left( \frac{\beta_4}{24}|u_{xx}|^2 + \frac{\beta_2}{2}|u_{x}|^2 + \frac{\gamma}{2} |u|^4 \right) dx.
\end{equation}
The energy $\mathcal{E}$ is invariant under the unitary rotation group $T_1(\theta)$, given by
\begin{equation}\label{eq:T1}
T_1(\theta) = \begin{pmatrix}
\cos(\theta) & -\sin(\theta) \\ 
\sin(\theta) & \cos(\theta)
\end{pmatrix},
\end{equation}
and the unitary translation group $T_2(s)$, given by $[T_2(s)]u(\cdot) = u(\cdot - s)$. These symmetries commute with $J$ and have infinitesimal generators
\begin{equation}\label{eq:Tgens}
T_1'(0) = \begin{pmatrix}
0 & -1 \\ 
1 & 0
\end{pmatrix} = -J, \qquad T_2'(0) = \partial_x.
\end{equation}
The corresponding conserved quantities are
\begin{equation}
Q_1 = -\frac{1}{2} \int_{\infty}^\infty |u|^2 dx,
\qquad Q_2 = \frac{1}{2} \int_{\infty}^\infty
\left( v_x w - w_x v \right) dx,
\end{equation}
which are the charge and momentum, respectively. We make the following hypothesis regarding the well-posedness of \cref{NLSHam}, which is the same as \cite[Assumption 1]{Grillakis1990}.

\begin{hypothesis}\label{hyp:wp}
For each initial condition $u_0$, there exists $T > 0$ depending only on $K$, where $\|u_0\| \leq K$, such that the PDE \cref{NLSHam} has a solution $u(t)$ on $[0, T]$ with $u(0) = u_0$.
\end{hypothesis}

Standing waves are solutions of the form $T_1(\omega t) u(x)$, where $u$ is independent of $t$. A standing wave solution satisfies the equation $\calE'(u) - \omega \calQ_1'(u) = 0$, which reduces to $\calE'(u) + \omega u = 0$ since $\calQ_1'(u) = -u$. For real-valued standing waves, this is equivalent to equation \cref{standingwavereal}. For the remainder of this paper, we will be concerned with solitary waves, which in this context are localized standing waves. The following theorem gives a criterion for the existence of real-valued, solitary wave solutions.

\begin{theorem}\label{theorem:solitonexist}
There exists a real-valued, symmetric, exponentially localized solitary wave solution $\phi(x; \omega) \in H^2(\R) \cap C^5(\R)$ to \cref{standingwavereal} if either (i) $\beta_2 \leq 0$ and $\omega > 0$, or (ii) $\beta_2 > 0$ and $\omega > \omega_c$, where 
\begin{equation}\label{omegac}
\omega_c = \frac{3}{2} \frac{\beta_2^2}{|\beta_4|}.
\end{equation}
The linearization $\calH = -\frac{\beta_4}{24} \partial_{xxxx} + \frac{\beta_2}{2} \partial_{xx} + \omega - 3 \gamma \phi^2$ of \cref{standingwavereal} about $\phi$ has exactly one negative eigenvalue with an even eigenfunction and a simple kernel with the odd eigenfunction $\partial_x \phi(x; \omega)$.
\begin{proof}
The existence result is similar to \cite[Theorem 2.1(i)]{Pelinovsky2007} (see in addition \cite[Theorem 1.1]{Bonheure2014} and \cite[Theorem 1.2]{Bonheure2018}). It follows from \cite{Groves1998} that a homoclinic orbit solution to the 4th order nonlinear ODE $r'''' + \mu r'' - cr = f(r, r', r'')$ exists when $c < 0$ and $\mu < 2 \sqrt{-c}$. (The required form of $f$ is given in \cite{Groves1998}). The homoclinic orbit is a critical point of an energy functional $\calJ(u)$, and the proof uses the mountain pass lemma and the concentration-compactness principle. Changing variables using \cref{rescaling}, equation \cref{eq:NLS4rescaled} is of this form with $c = -1$ and $\mu = 2 \sigma$, thus a solution $U_0(x)$ exists for $\sigma < 1$. It follows from the proof of \cite[Theorem 2.1(ii)]{Pelinovsky2007} that the linearization $\partial_X^4 + 2 \sigma \partial_X^2 + I - 3 U_0^2$ of \cref{eq:NLS4rescaled} about $U_0$ has exactly one negative eigenvalue with an even eigenfunction and a simple kernel with the odd eigenfunction $U_0'$. Exponential localization follows from the stable manifold theorem. The result, and the specific value of $\omega_c$, follow upon reversing the change of variables \cref{rescaling}.
\end{proof}
\end{theorem}

\begin{remark}\label{remark:soliton}
The set of $(\omega, \beta_2)$ for which solutions to \cref{standingwavereal} exist can be divided into two disjoint regions. For $\beta_2 < 0$ and $0 < \omega < \omega_c$, the primary solitary wave solutions have exponentially decaying, non-oscillatory tails. For $\omega > \omega_c$ and all $\beta_2$, the primary solitary wave solutions have tails which are exponentially decaying and oscillatory. See \cite[Figure 2(a)]{Tam2020} (the frequency $\omega$ is denoted by $\mu$ in that paper). In addition, we note that $\omega_c = 0$ for PQS, thus PQS exist for all $\omega > 0$.
\end{remark}

\noi We make the following standard smoothness assumption (see, for example, \cite[Assumption 2]{Grillakis1987}) concerning the solutions $\phi(x; \omega)$ to \cref{standingwavereal}.

\begin{hypothesis}\label{hyp:smoothmap}
The map $\omega \mapsto \phi(x; \omega)$ from $\calI$ to $H^2(\R)$ is $C^1$, where $\calI$ is the interval for which the primary pulse solution $\phi(x; \omega)$ exists.
\end{hypothesis}

Let $\beta_4 < 0$ and $\beta_2 \in \R$, and choose $\omega > 0$ such that the primary pulse solution $\phi(x) = \phi(x; \omega)$ exists by \cref{theorem:solitonexist}. From this point forward, we will suppress the dependence on $\omega$ for simplicity of notation. The linearization of the PDE \cref{NLS4} about $\phi$ is the linear operator $J \calL(\phi)$, where $\calL(\phi): H^4(\R) \subset L^2(\R) \mapsto L^2(\R)$ is given by
\begin{align}\label{defLphi}
\calL(\phi) = 
\begin{pmatrix}
\calL^+(\phi) & 0 \\
0 & \calL^-(\phi)
\end{pmatrix},
\end{align}
and
\begin{align*}
\calL^-(\phi) &= -\frac{\beta_4}{24} \partial_{xxxx} + \frac{\beta_2}{2} \partial_{xx} + \omega - \gamma \phi^2 \\
\calL^+(\phi) &= -\frac{\beta_4}{24} \partial_{xxxx} + \frac{\beta_2}{2} \partial_{xx} + \omega - 3 \gamma \phi^2.
\end{align*}

Both $\calL^-(\phi)$ and $\calL^+(\phi)$ are self-adjoint, thus their spectrum is real. Since $\calL^+(\phi)$ is the same as the linear operator $\calH$ from \cref{theorem:solitonexist}, $\calL^+(\phi)$ has a single negative eigenvalue and a simple kernel with eigenfunction $\partial_x \phi$. It is also straightforward to verify that $\calL^-(\phi) \phi = 0$. We make the following hypothesis concerning the spectrum of $\calL^-(\phi)$.
\begin{hypothesis}\label{hyp:Lminusspec}
The operator $\calL^-(\phi)$ has no negative eigenvalue and has a simple kernel with eigenfunction $\phi$.
\end{hypothesis}
This hypothesis is proved in \cite{Natali2015} for the specific case of one of the Karlsson-H{\"o}{\"o}k solitons \cite{KarllsonHook}. It follows from \cref{theorem:solitonexist}, \cref{hyp:Lminusspec}, and the fact that $\calL(\phi)$ is diagonal and $J$ is invertible that the kernel of $J \calL(\phi)$ has geometric multiplicity 2. We can additionally verify that $\calL^+(\phi)(-\partial_\omega \phi) = \phi$. Since $\calL^-(\phi)$ is self-adjoint and $\phi \perp \ker \calL^-(\phi)$ by \cref{hyp:Lminusspec}, there exists a function $z$ such that $\calL^-(\phi) z = \partial_x \phi$ by the Fredholm alternative. For the classical NLS equation, $z = \frac{1}{2 \beta_2} x \phi$. We then have
\begin{equation}\label{Lphikernel}
\begin{aligned}
J \calL(\phi)\begin{pmatrix}0 \\ \phi \end{pmatrix} &= 0, \quad
J \calL(\phi)\begin{pmatrix} \partial_\omega \phi \\ 0 \end{pmatrix} = \begin{pmatrix}0 \\ \phi \end{pmatrix}, \\
J \calL(\phi)\begin{pmatrix}\partial_x\phi \\ 0 \end{pmatrix} &= 0, \quad
J \calL(\phi)\begin{pmatrix} 0 \\ z \end{pmatrix} = \begin{pmatrix}\partial_x\phi \\ 0 \end{pmatrix},
\end{aligned}
\end{equation}
thus the kernel of $J \calL(\phi)$ has algebraic multiplicity at least 4. To show that there are no more generalized eigenfunctions in the kernel of $J \calL(\phi)$, we will need one additional hypothesis. 
\begin{hypothesis}\label{hyp:Mcond}
For each $\omega$ such that a primary pulse solution $\phi(x; \omega)$ exists
\begin{align}
M &:= \partial_\omega \frac{1}{2} \| \phi \|_{L^2(\R)}^2
= \int_{-\infty}^\infty \phi(x) \partial_\omega \phi(x) dx > 0, \label{eq:M} \\
\tilde{M} &:= \langle \partial_x \phi, z \rangle_{L^2(\R)} = \int_{-\infty}^\infty z(x) \partial_x \phi(x) dx > 0. \label{eq:tildeM}
\end{align}
\end{hypothesis}
The condition \cref{eq:M} is the Vakhitov–Kolokolov stability criterion \cite{Vakhitov1973}. The two conditions in \cref{hyp:Mcond}, together with \cref{hyp:Lminusspec}, are sufficient to prove orbital stability of the solitary wave solution $\phi$, as shown in the following lemma.
\begin{lemma}\label{lemma:stability}
Assume \cref{hyp:smoothmap}, \cref{hyp:Lminusspec}, and \cref{hyp:Mcond}. Then the primary solitary wave solution $\phi(x; \omega)$ is orbitally stable.
\begin{proof}
This follows from \cite[Section 5.2.2]{Kapitula2013}, which is an extension of the techniques in \cite{Grillakis1990}. Since $\calL^+(\phi)$ has a single negative eigenvalue by \cref{theorem:solitonexist}, $\calL^-(\phi)$ has no negative eigenvalues by \cref{hyp:Lminusspec}, $J$ is invertible, and $\calL(\phi) = \text{diag}(\calL^+(\phi), \calL^-(\phi))$, $\calL(\phi)$ has a single negative eigenvalue. Next, we form the symmetric matrix $D$ \cite[(5.2.53)]{Kapitula2013}, defined by 
$D_{ij} = \langle \calL(\phi)^{-1} s_i, s_j \rangle$, where $s_j = J^{-1} T_j'(0) \phi$ for $j = 1, 2$. Using \cref{eq:Tgens}, $s_1 = (-\phi, 0)^T$ and $s_2 = (0, \partial_x \phi)$, thus $\calL(\phi)^{-1} s_1 = (\partial_\omega \phi, 0)^T$ and $\calL(\phi)^{-1} s_2 = (0, z)^T$. It follows that $D = \text{diag}(-M, \tilde{M})$. By \cref{hyp:Mcond}, $D$ has a single negative eigenvalue. Since $n(\calL(\phi)) = n(D)$, i.e. $\calL(\phi)$ and $D$ have the same number of negative eigenvalues, the primary solitary wave solution $\phi$ is orbitally stable by \cite[Theorem 5.2.11]{Kapitula2013}.
\end{proof}
\end{lemma}
\begin{remark}We can equivalently assume \cref{hyp:Lminusspec}, the Vakhitov–Kolokolov stability criterion $M > 0$, $\tilde{M} \neq 0$, and the orbital stability of the primary solitary wave. If $\tilde{M} < 0$, then it then follows from the Jones–Grillakis instability index \cite{Grillakis1988}, an extension of Hamiltionian-Krein index theory \cite[Section 7.1]{Kapitula2013}, that there is an eigenvalue with positive real part \cite[Theorem 7.1.16]{Kapitula2013}, which contradicts orbital stability. Thus we must have $\tilde{M} > 0$.
\end{remark}
It follows from \cref{hyp:Mcond} that the algebraic multiplicity of the kernel of $J \calL(\phi)$ is exactly 4.
\begin{lemma}\label{lemma:kernelL}
Assume \cref{hyp:Lminusspec} and \cref{hyp:Mcond}. Then the kernel of $J \calL(\phi)$ has geometric multiplicity 2 and algebraic multiplicity 4.
\begin{proof}
We showed above that $\ker J \calL(\phi)$ has geometric multiplicity 2 and algebraic multiplicity at least 4. All that remains is to show that there can be no other generalized eigenfunctions. If $J \calL(\phi) u = (\partial_\omega \phi, 0)^T$ has a solution, then $\calL^- w = \partial_\omega \phi$ has a solution, which implies $\partial_\omega \phi \perp \ker(\calL^-(\phi)) = \spn\{\phi\}$ by the Fredholm alternative and \cref{hyp:Lminusspec}, contradicting \cref{eq:M}. If $J \calL(\phi) u = (0, z)^T$ has a solution, then $\calL^+ v = -z$ has a solution, which implies $z \perp \ker(\calL^+(\phi)) = \spn\{\partial_x \phi\}$ by the Fredholm alternative and \cref{theorem:solitonexist}, contradicting \cref{eq:tildeM}.
\end{proof}
\end{lemma}

We now discuss the spectrum of $J \calL(\phi)$. The spectrum of $J \calL(\phi)$ can be divided into two disjoint sets: the essential spectrum is the set of $\lambda \in \C$ for which $J \calL(\phi) - \lambda \calI$ is not Fredholm \cite[Section 3.1]{Kapitula2013}, and the point spectrum is the set of $\lambda \in \C$ for which $\ker( J \calL(\phi) - \lambda \calI)$ is nontrivial. To find the essential spectrum, which depends only on the background state and is independent of the solution $\phi$ we are linearizing about, $J \calL(\phi)$ is exponentially asymptotic to the linear operator $J \calL(0)$, given by
\begin{align}\label{defL0}
J \calL(0) = 
\begin{pmatrix}
0 & \calL_0 \\
-\calL_0 & 0
\end{pmatrix}, \quad
\calL_0 = -\frac{\beta_4}{24} \partial_{xxxx} + \frac{\beta_2}{2} \partial_{xx} + \omega.
\end{align}
The eigenvalue problem $J \calL(0) v = \lambda v$ is equivalent to $(\calL_0^2 + \lambda^2)p = 0$. By \cite[Theorem 3.1.13]{Kapitula2013}, the essential spectrum is given by the curves
\begin{align*}
\left[ -\frac{\beta_4}{24} (ik)^4 + \frac{\beta_2}{2}(ik)^2 + \omega \right]^2 + \lambda^2 &= 0 && k \in \R,
\end{align*}
from which it follows that
\begin{align*}
\sigma_{\text{ess}} = \left\{ \pm i \left( -\frac{\beta_4}{24}k^4 - \frac{\beta_2}{2}k^2 + \omega \right) : k \in \R \right\}.
\end{align*}
If $\beta_4 < 0$ and $\beta_2 \leq 0$, the essential spectrum is given by 
\begin{equation}\label{PQSessspec}
\sigma_{\text{ess}} = \{ k i : k \in \R, |k| \geq \omega \},
\end{equation}
which is purely imaginary, bounded away from the origin, and independent of $\beta_4$ and $\beta_2$. In particular, this is the case for PQS. If $\beta_4 < 0$, $\beta_2 > 0$, and $\omega > \omega_c$, the essential spectrum is given by 
\begin{equation}\label{essspec2}
\sigma_{\text{ess}} = \{ k i : k \in \R, |k| \geq \omega - \omega_c \},
\end{equation}
which is also purely imaginary and bounded away from the origin, but does depend on $\beta_4$ and $\beta_2$ via $\omega_c$.

We now turn to the point spectrum of $J \calL(\phi)$. By \cref{lemma:kernelL}, the kernel of $J \calL(\phi)$ is given by \cref{Lphikernel}. Since the primary solitary wave is orbitally stable by \cref{lemma:stability}, no element of the spectrum of $J \calL(\phi)$ can have positive real part. Since the PDE \cref{NLS4} is Hamiltonian, all elements of the spectrum of $J \calL(\phi)$ must come in quartets $\pm \alpha \pm \beta i$ \cite[Proposition 7.0.1]{Kapitula2013}, thus the spectrum of $J \calL(\phi)$ is contained in the imaginary axis. For PQS, there is an additional pair of imaginary eigenvalues located right before the essential spectrum boundary (approximately $\pm 0.9972 \omega i$ for $\beta_4 = -1$), which corresponds to an internal mode of the solitary wave \cite{Tam2019}. For $\beta_2 \neq 0$, there can be multiple pairs of internal mode eigenvalues (an example of two pairs of internal mode eigenvalues is shown in \cite[Figure 9]{Tam2020}). By \cref{lemma:stability}, these internal mode eigenvalues must be purely imaginary.

\section{Existence of multi-pulse solitary waves}

A multi-pulse is a multi-modal solitary wave resembling multiple, well-separated copies of the primary solitary wave. To prove the existence of multi-pulse solutions to \cref{standingwavereal}, we will reframe the problem using a spatial dynamics approach. From this perspective, the primary solitary wave is a homoclinic orbit connecting the unstable and stable manifolds of a saddle equilibrium at the origin. A multi-pulse is a multi-loop homoclinic orbit which remains close to the primary homoclinic orbit. Letting $U = (u_1, u_2, u_3, u_4) = (u, \partial_x u, \partial_x^2 u, \frac{\beta_4}{24} \partial_x^3 u)$, we rewrite equation \cref{standingwavereal} as the first order system
\begin{equation}\label{Fsystem}
U' = F(U) = \begin{pmatrix}
u_2 \\ u_3 \\ \frac{24}{\beta_4} u_4 \\ \omega u_1 - \gamma u_1^3
\end{pmatrix}.
\end{equation}
This system has a conserved quantity
\begin{equation}\label{FsystemH}
H(u_1, u_2, u_3, u_4) = -u_4 u_2 - \frac{1}{2} u_3^2 + \frac{\beta_2}{4}u_2^2 - \frac{\gamma}{4} u_1^4 + \frac{1}{2}\omega u_1^2,
\end{equation}
which we obtain by multiplying \cref{standingwavereal} by $u_x$ and integrating once. $F(0) = 0$, and the characteristic polynomial of $DF(0)$ is
\[
p(t) = t^4 - 12\frac{\beta_2}{\beta_4} t^2 - \frac{24}{\beta_4}\omega,
\]
which has a quartet of complex eigenvalues $\pm a \pm b i$ when $\omega > \omega_c$, where $\omega_c$ is defined in \cref{theorem:solitonexist}. For $\omega > \omega_c$, $U = 0$ is a hyperbolic saddle equilibrium of \cref{Fsystem} with two-dimensional stable and unstable manifolds which intersect to form a homoclinic orbit. The exponentially localized primary pulse solution corresponding to this homoclinic orbit will have oscillatory tails, with the frequency of oscillations approximately equal to $b$. We have the following result concerning the existence of multi-pulse solutions, which follows immediately from \cite[Theorem~3.6]{SandstedeStrut}. 

\begin{theorem}\label{theorem:multiexist}
Assume \cref{hyp:wp} and \cref{hyp:smoothmap}, and fix $\beta_4 < 0$ and $\omega > \omega_c$, where $\omega_c$ is defined in \cref{theorem:solitonexist}. Let $\phi(x)$ be the real-valued, symmetric, exponentially localized primary pulse solution to \cref{standingwavereal} from \cref{theorem:solitonexist}, and let $U(x) = (\phi(x), \partial_x \phi(x), \partial_x^2 \phi(x), \partial_x^3 \phi(x))$ be the corresponding homoclinic orbit solution to \cref{Fsystem}. Let $\pm a \pm bi$ be the eigenvalues of $DF(0)$, with $a > 0$ and $b > 0$. Then for any 
\begin{compactenum}[(i)]
\item $n \geq 2$
\item Sequence of nonnegative integers $\{ k_1, \dots, k_{n-1} \}$, with at least one of the $k_j \in \{0, 1 \}$
\item Sequence of phase parameters $\{ \theta_1, \dots, \theta_n \} \in \{-1, 1 \}^n$, with $\theta_1 = 1$
\end{compactenum}
there exists a nonnegative integer $m_0$ such that for any integer $m$ with $m \geq m_0$, there exists a unique $n-$modal solution $U_n(x)$ to \cref{Fsystem} which is defined piecewise via
\begin{equation}\label{Unpiecewise}
U_n\left( x + 2 \sum_{k=1}^{i-1} X_k \right) = \begin{cases} 
\theta_i U(x) + \tilde{U}_i^-(x) & x \in [-X_{i-1}, 0] \\
\theta_i U(x) + \tilde{U}_i^-(x) & x \in [0, X_i]
\end{cases}
\end{equation}
for $i = 1, \dots, n$, where $X_0 = X_n = \infty$. Uniqueness is up to translation and multiplication by $T(\theta)$. The distances between consecutive peaks are given by $2 X_i$, where
\begin{equation}\label{pulsedistances}
X_i \approx \frac{\pi}{b}(2 m + k_i) + \tilde{X},
\end{equation}
and $\tilde{X}$ is a constant. In addition, we have the estimates
\begin{equation}\label{Unestimates}
\begin{aligned}
\|\tilde{U}_i^\pm\|_\infty &\leq C e^{-a X_{\mathrm{min}}} \\
\tilde{U}_i^+(X_i) &= \theta_{i+1} U(-X_i) + \mathcal{O}(e^{-2 a X_{\min}}) \\
\tilde{U}_{i+1}^-(-X_i) &= \theta_i U(X_i) + \mathcal{O}(e^{-2 a X_{\min}}),
\end{aligned}
\end{equation}
where $X_{\mathrm{min}} = \min \{ X_1, \dots X_{n-1} \}$, which hold as well for all derivatives with respect to $x$.
\begin{proof}
Since the spectrum of $DF(0)$ is a quartet of eigenvalues $\pm a \pm b i$ for $\omega > \omega_c$, equation \cref{Fsystem} has a conserved quantity \cref{FsystemH}, and the Melnikov integral $M = \int_{-\infty}^\infty \phi_x^2 dx$ is positive, the result follows from \cite[Theorem~3.6]{SandstedeStrut}, with the straightforward modification that the multi-pulse is constructed from copies of $U(x)$ and $-U(x)$. The estimates \cref{Unestimates} follow from \cite{Sandstede1993, Sandstede1998}.
\end{proof}
\end{theorem}

\section{Spectrum of multi-pulse solitary waves}\label{sec:multieig}

Let $U(x) = (\phi(x), \partial_x \phi(x), \partial_x^2 \phi(x), \frac{\beta_4}{24} \partial_x^3 \phi(x))$ be the primary homoclinic orbit corresponding to the primary pulse $\phi(x)$, and let $U_n = (\phi_n(x), \partial_x \phi_n(x), \partial_x^2 \phi_n(x), \frac{\beta_4}{24} \partial_x^3 \phi_n(x))$ be a multi-loop homoclinic orbit solution to \cref{Fsystem} constructed according to \cref{theorem:multiexist}. The first component $\phi_n$ of $U_n$ is a multi-pulse solitary wave solution to \cref{standingwavereal}. As in \cite{Sandstede1998,Manukian}, we will locate the eigenvalues near the origin of the linearized operator $J \calL(\phi_n)$. As shown above in \cref{Lphikernel}, $J \calL(\phi)$ has two eigenfunctions in the kernel from the gauge symmetry and translational invariance, as well as two additional generalized eigenfunctions. $J \calL(\phi_n)$ will also have two eigenvalues in the kernel from the same symmetries. It remains to locate the interaction eigenvalues, which arise from nonlinear interactions between the tails of neighboring pulses in the multi-pulse structure. Once again using a spatial dynamics approach, we rewrite the eigenvalue problem $J \calL(\phi_n)v = \lambda v$ as the first order system
\begin{equation}\label{multieig}
V'(x) = K(\phi_n)V(x) + \lambda B_1 V(x),
\end{equation}
where
\begin{align*}
K(\phi_n) &= 
\begin{pmatrix}K^+(\phi_n) & 0 \\ 0 & K^-(\phi_n) \end{pmatrix}, \quad
B_1 = \begin{pmatrix}0 & B \\ -B & 0\end{pmatrix}, \\
K^-(\phi_n) &= \begin{pmatrix}
0 & 1 & 0 & 0 \\
0 & 0 & 1 & 0 \\
0 & 0 & 0 & \frac{24}{\beta_4} \\
\omega - \gamma \phi_n^2 & 0 & \frac{\beta_2}{2} & 0
\end{pmatrix},
K^+(\phi) = \begin{pmatrix}
0 & 1 & 0 & 0 \\
0 & 0 & 1 & 0 \\
0 & 0 & 0 & \frac{24}{\beta_4} \\
\omega - 3 \gamma \phi_n^2 & 0 & \frac{\beta_2}{2} & 0
\end{pmatrix},
B = \begin{pmatrix}
0 & 0 & 0 & 0 \\
0 & 0 & 0 & 0 \\
0 & 0 & 0 & 0 \\
1 & 0 & 0 & 0
\end{pmatrix}.
\end{align*}
The associated variational equation 
\begin{align}
V'(x) = K(\phi)V(x) \label{vareq}
\end{align}
has two linearly independent, exponentially decaying solutions $\tilde{Q}(x) = (U'(x), 0)^T$ and $Q(x) = (0, U(x))^T$. The corresponding adjoint variational equation
\begin{align}
W'(x) = -K(\phi)^*W(x)\label{adjvareq}
\end{align}
has two linearly independent, exponentially decaying solutions $\tilde{Q}^*(x) = (\Psi'(x), 0)^T$ and $Q^*(x) = (0, \Psi(x) )^T$, where
\begin{equation}\label{defPsi}
\Psi(x) =
\left( -\frac{\beta_4}{24} \partial_x^3 \phi(x) + \frac{\beta_2}{2} \partial_x \phi(x),
\frac{\beta_4}{24} \partial_x^2 \phi(x) - \frac{\beta_2}{2} \phi(x),
- \frac{\beta_4}{24} \partial_x \phi(x), \phi(x) \right).
\end{equation}

The following theorem, which is analogous to the results in \cite[Section 3.4]{Manukian} and involves the same tri-diagonal matrix as in \cite[Theorem 2]{Sandstede1998}, reduces the problem of locating the eigenvalues of $J \calL(\phi_n)$ in a ball around the origin in the complex plane to finding the determinant of a $2n\times 2n$ matrix which is, to leading order, block diagonal. The proof is given in \cref{sec:blockmatrixproof}.

\begin{theorem}\label{th:blockmatrix}
Assume \cref{hyp:wp}, \cref{hyp:smoothmap}, \cref{hyp:Lminusspec}, and \cref{hyp:Mcond}. Let $U(x)$ be the primary homoclinic orbit from \cref{theorem:solitonexist}, and let $U_n(x)$ be an $n$-pulse solution constructed according to \cref{theorem:multiexist} with phase parameters $\{ \theta_1, \dots, \theta_n \}$ and pulse distances $X_1, \dots, X_{n-1}$. Let $\pm a \pm bi$ be the eigenvalues of $DF(0)$, with $a > 0$ and $b > 0$. Then there exists $\delta > 0$ with the following property. There exists a bounded, nonzero solution $V(x)$ of \cref{multieig} for $|\lambda| < \delta$ if and only if
\begin{equation}\label{blockmatrixcond}
E(\lambda) = \det S(\lambda) = 0,
\end{equation}
where $S(\lambda)$ is the $2n \times 2n$ block matrix
\begin{equation}\label{blockeq}
S(\lambda) = 
\begin{pmatrix}
A + \lambda^2 M I & 0 \\
0 & (a^2 + b^2) A - \lambda^2 \tilde{M} I
\end{pmatrix} + R(\lambda).
\end{equation}
The tri-diagonal matrix $A$ is defined by
\begin{equation*}
A = \begin{pmatrix}
-a_1 & a_1 \\
a_1 & -a_1 - a_2 &  a_2 \\
& a_2 & -a_2 - a_3 &  a_3 \\
& & \ddots & \ddots \\
& & &  a_{n-1} & -a_{n-1} \\
\end{pmatrix},
\end{equation*}
where
\begin{align*}
a_i &= \theta_i \theta_{i+1} \langle \Psi(X_i), U(-X_i) \rangle,
\end{align*}
$\Psi(x)$ is defined by \cref{defPsi}, and the constants $M$ and $\tilde{M}$ are defined in \cref{hyp:Mcond}. The remainder term $R(\lambda)$ is analytic in $\lambda$ and has uniform bound
\[
|R(\lambda)| \leq C\left( |\lambda|(|\lambda| + e^{-\alpha X_{\min}})^2 + e^{-(2 \alpha + \gamma)X_{\min} }) \right),
\]
where $\gamma > 0$.
\end{theorem}

The interaction eigenvalues can be computed using the eigenvalues of the matrix $A$ from \cref{th:blockmatrix}. It follows that all multi-pulse solutions have a positive, real eigenvalue and thus are unstable. 

\begin{corollary}\label{corr:multiunstable}
Assume the same hypotheses as in \cref{th:blockmatrix}. Let $U_n(x)$ be a n-pulse constructed using \cref{theorem:multiexist}. Then there are $2(n-1)$ pairs of interaction eigenvalues $\lambda_1, \dots \lambda_{n-1}$ and $\tilde{\lambda}_1, \dots \tilde{\lambda}_{n-1}$, given by
\begin{equation}\label{inteigs}
\begin{aligned}
\lambda_i &= \sqrt{\frac{\mu_i}{M}} + \mathcal{O}\left( e^{-(2 \alpha + \gamma)X_{\min} } \right) && i = 1, \dots, n-1 \\
\tilde{\lambda}_i &= \sqrt{-(a^2 + b^2) \frac{\mu_i}{\tilde{M}}} + \mathcal{O}\left( e^{-(2 \alpha + \gamma)X_{\min} } \right) && i = 1, \dots, n-1,
\end{aligned}
\end{equation}
where $\{ \mu_1,\dots,\mu_{n-1}, 0\}$ are the real, distinct eigenvalues of $A$. One of each pair $\{\lambda_i, \tilde{\lambda}_i\}$ is real and the other is purely imaginary, thus there are $n-1$ positive real eigenvalues. There is also an eigenvalue with algebraic multiplicity 4 and geometric multiplicity 2 at the origin.
\end{corollary}

Intuitively, the instability result is a consequence of the opposite signs of the $\lambda^2$ terms on the diagonal in \cref{blockeq}. This follows directly from the Hamiltonian structure of the PDE \cref{NLSHam}, in particular the opposite signs of the off-diagonal terms in the symplectic matrix $J$. Finally, we compute the interaction eigenvalues of a 2-pulse solution $U_2(x)$ explicitly.

\begin{corollary}\label{corr:2pstab}
Assume the same hypotheses as in \cref{th:blockmatrix}. Let $U_2(x)$ be a 2-pulse constructed using \cref{theorem:multiexist} with pulse distance $X_1$ and phase parameters $\theta_1, \theta_2$. Then there are four interaction eigenvalues associated with $U_2(x)$, which are, to leading order, given by
\begin{equation}\label{inteigpred}
\begin{aligned}
\lambda &= \pm \sqrt{ \frac{2 a_1}{M} }, \quad
\tilde{\lambda} = \pm \sqrt{ \frac{-2 (a^2 + b^2) a_1}{\tilde{M}} },
\end{aligned}
\end{equation}
where $a_1 = \theta_1 \theta_2 \langle \Psi(X_1), U(-X_1) \rangle$. One pair is real and one pair is purely imaginary. There is also an eigenvalue with algebraic multiplicity 4 and geometric multiplicity 2 at the origin.
\end{corollary}

\begin{remark}In addition, the internal mode eigenvalues of the primary pulse will duplicate as pulses are added to the multi-pulse structure. For example, for the pure quartic solitary wave, the 2-pulse will have two pairs of internal mode eigenvalues. Since the 2-pulse is unstable, these internal mode eigenvalues have no additional effect on stability.
\end{remark}

\section{Numerical results}

\subsection{Primary pulse}

\begin{figure}[H]
\centering
\begin{tabular}{cc}
\includegraphics[width=8cm]{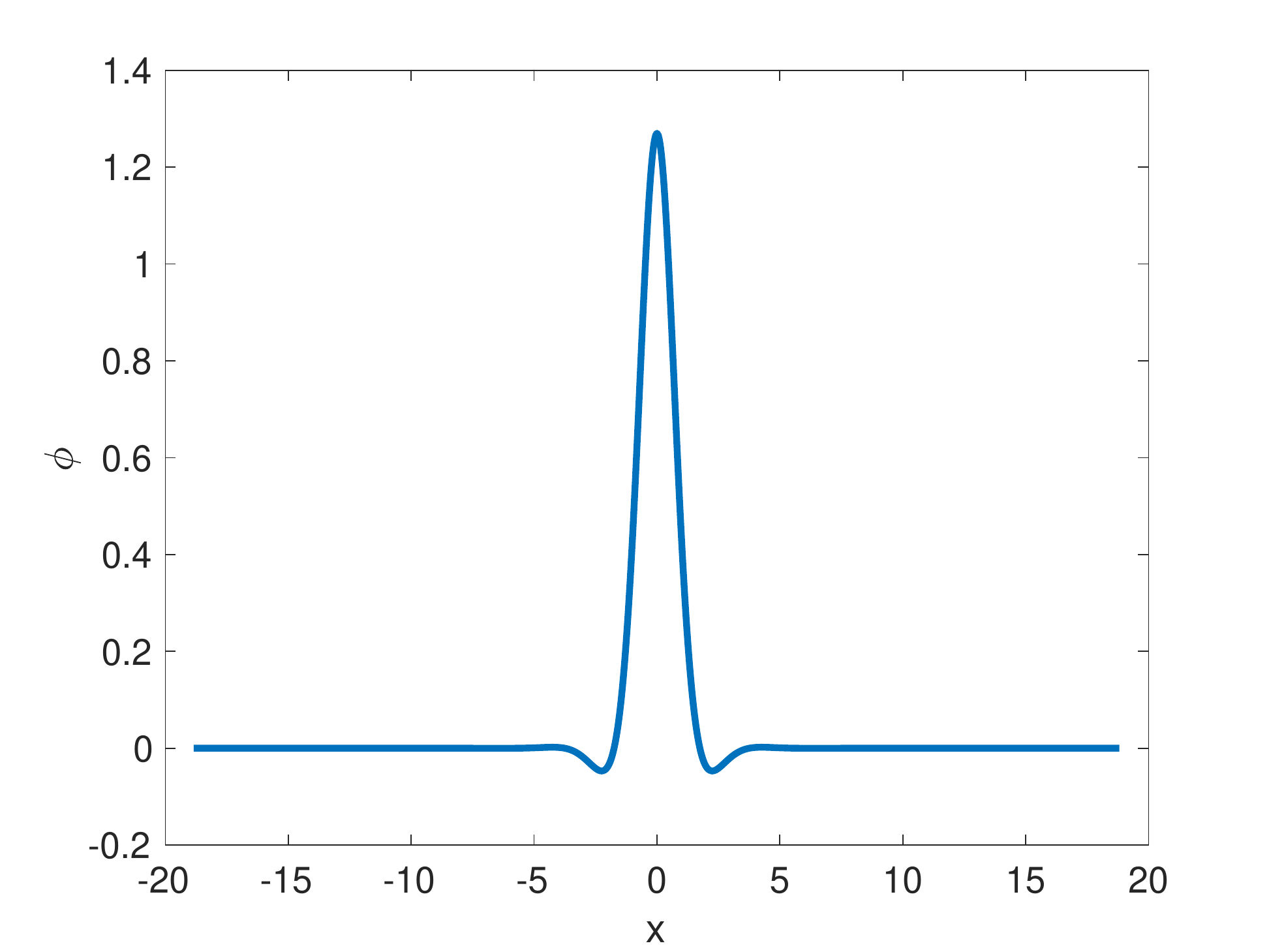} &
\includegraphics[width=8cm]{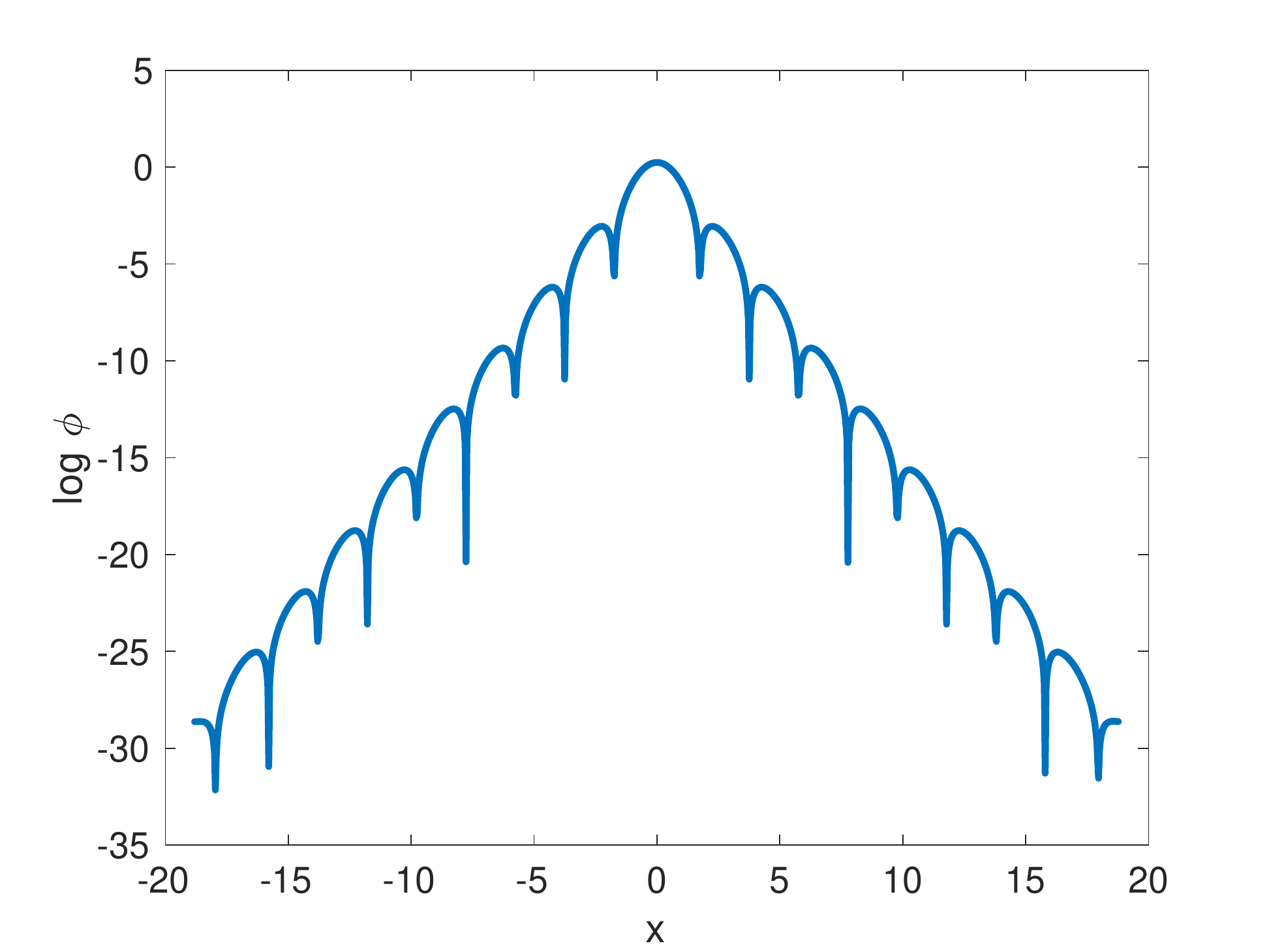}
\end{tabular}
\caption{Pure quartic solitary wave solution $\phi(x)$ to \cref{standingwavereal} with $\beta_2 = 0$, $\beta_4 = -1$, $\omega = 1$, and $\gamma = 1$. (left panel). Plot of $\log \phi(x)$ vs $x$ (right panel) showing exponentially-decaying oscillatory tails. Spatial discretization is a uniform grid with $N = 1024$ grid points, and we use periodic boundary conditions. }
\label{fig:PQS}
\end{figure}

To construct the primary pulse solution $\phi(x)$, we start with the known solitary wave solution for NLS and gradually modify the parameters $\beta_2$ and $\beta_4$, solving for the new solitary wave solution at each step using a Newton conjugate-gradient method \cite[Chapter 7.2.4]{YangCh7} implemented in MATLAB. To obtain the pure quartic solitary wave for $\beta_4 = -1$ (\cref{fig:PQS}), we perform this procedure along the line segment connecting $(\beta_2, \beta_4) = (-2, 0)$ and $(\beta_2, \beta_4) = (0, -1)$.

\begin{figure}[H]
\centering
\begin{tabular}{c}
\includegraphics[width=8cm]{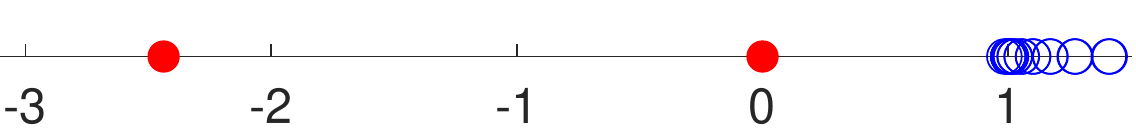}\\
\includegraphics[width=8cm]{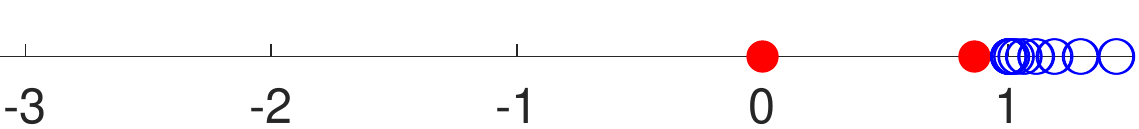}
\end{tabular}
\caption{Point spectrum (red dots) and essential spectrum (blue open circles) of $\calL^+(\phi)$ (top) and $\calL^-(\phi)$ (bottom) for PQS solution to \cref{standingwavereal}. $\beta_2 = 0$, $\beta_4 = -1$, $\omega = 1$, and $\gamma = 1$. For these parameters, the point spectrum of $\calL^+(\phi)$ contains an additional internal mode eigenvalue just to the left of the essential spectrum border, which is not shown.}
\label{fig:Lpmspec}
\end{figure}

The spectrum of the linear operators $\calL^+(\phi)$ and $\calL^-(\phi)$ is shown in \cref{fig:Lpmspec}, which confirms the results of \cref{theorem:solitonexist} and validates \cref{hyp:Lminusspec}. The point spectrum of $\calL^-(\phi)$ contains an additional positive eigenvalue which does not affect the results above. In addition, we can verify that $M > 0$ and $\tilde{M} > 0$ from \cref{hyp:Mcond}, from which it follows by \cref{lemma:stability} that the primary solitary wave $\phi$ is orbitally stable.

\begin{figure}[H]
\centering
\begin{tabular}{cc}
\includegraphics[width=8cm]{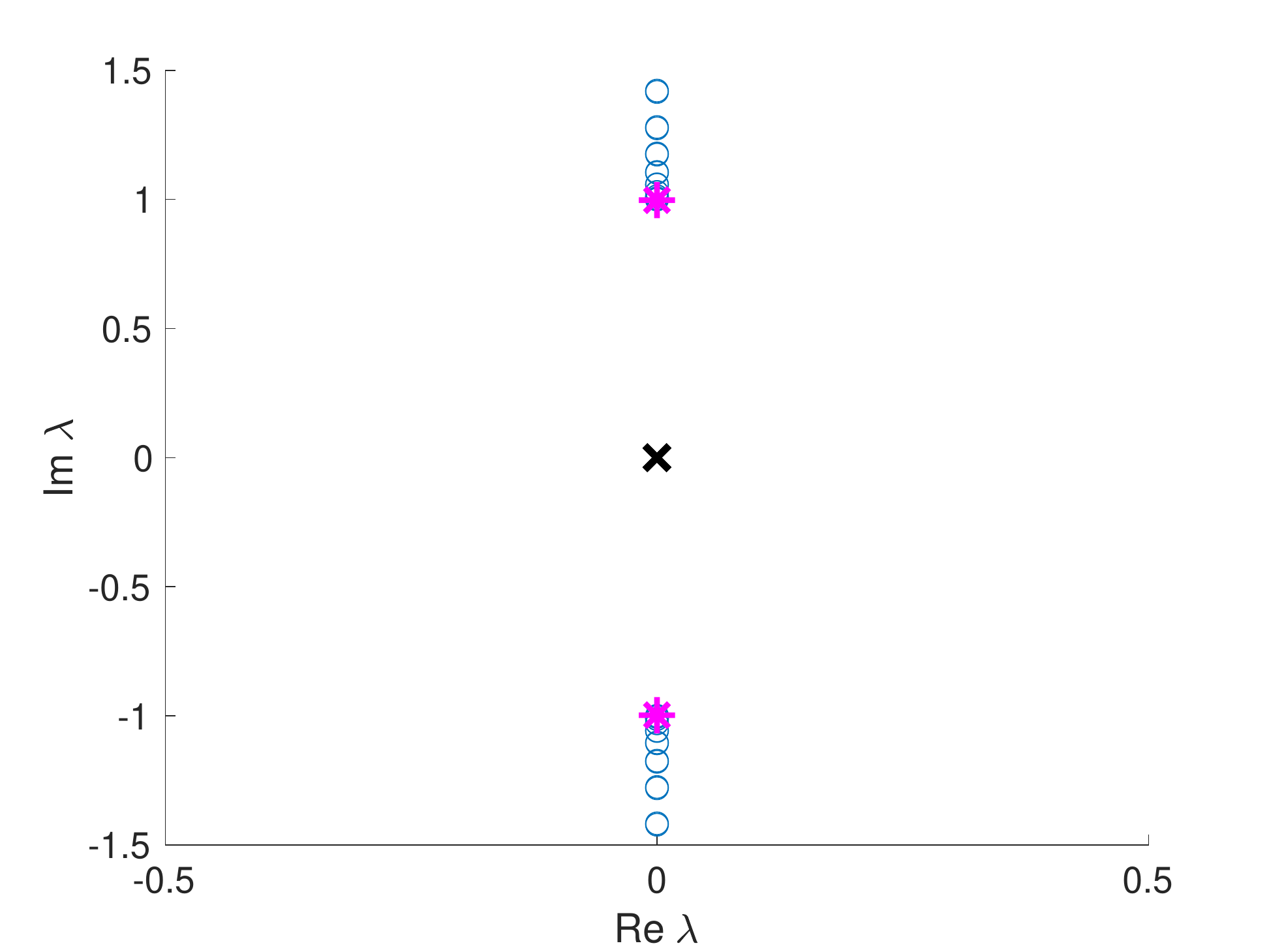} &
\includegraphics[width=8cm]{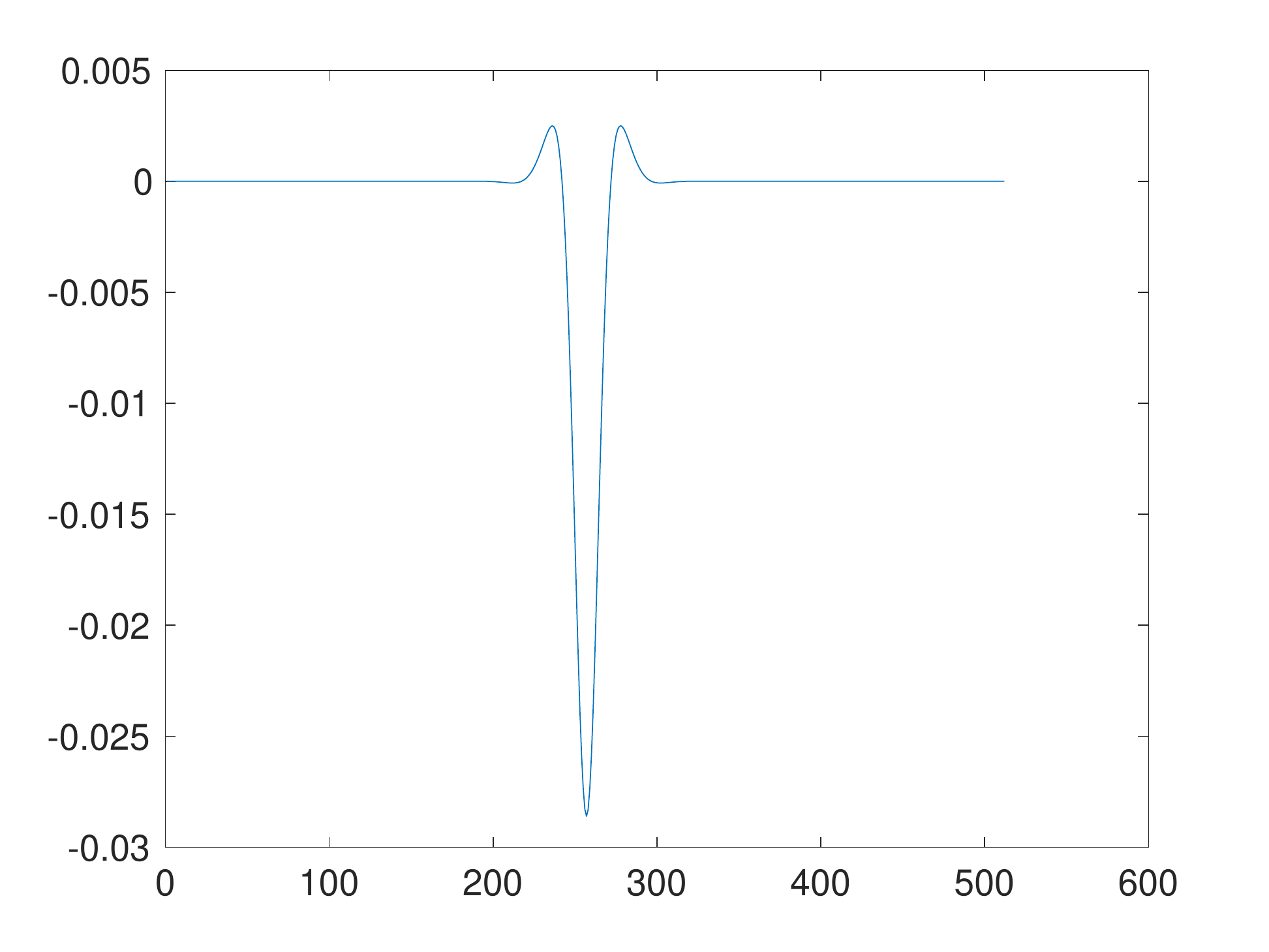}
\end{tabular}
\caption{Spectrum of pure quartic solitary wave (left panel), $\beta_2 = 0$, $\beta_4 = -1$, $\omega = 1$, and $\gamma = 1$. Spectrum comprises essential spectrum (blue open circles), internal mode eigenvalues (magenta stars), and kernel eigenvalue with geometric multiplicity 2 and algebraic multiplicity 4 (black X). Eigenfunction corresponding to internal mode eigenvalue (right panel).}
\label{fig:PQSspec}
\end{figure}

To determine the spectrum of the linearization about the primary pulse, we construct the linear operator $J \calL(\phi)$ using Fourier spectral differentiation matrices with periodic boundary conditions and compute the eigenvalues using Matlab's eigenvalue solver \texttt{eig} (\cref{fig:PQSspec}, left panel). Since the spectral problem is posed on a periodic domain, the essential spectrum is discrete. It is a finite set of points in this case since spatial discretization approximates the eigenvalue problem \cref{multieig} with an $2N \times 2N$ matrix equation, where $N$ is the number of grid points. We also note the presence of a pair of internal mode eigenvalues on the imaginary axis.  

\subsection{Construction of multi-pulses}

\begin{figure}[H]
\centering
\begin{tabular}{cc}
\includegraphics[width=8cm]{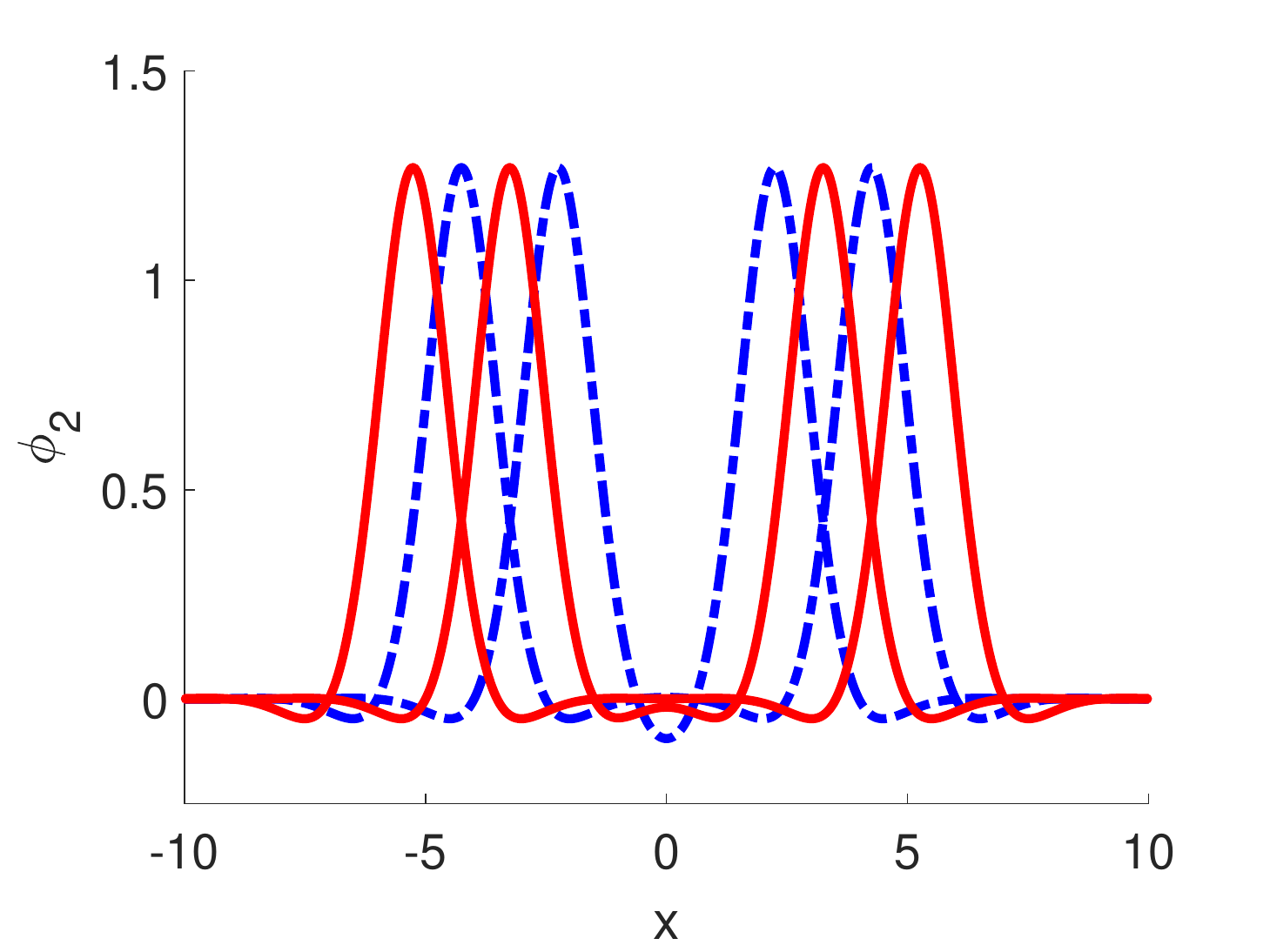} &
\includegraphics[width=8cm]{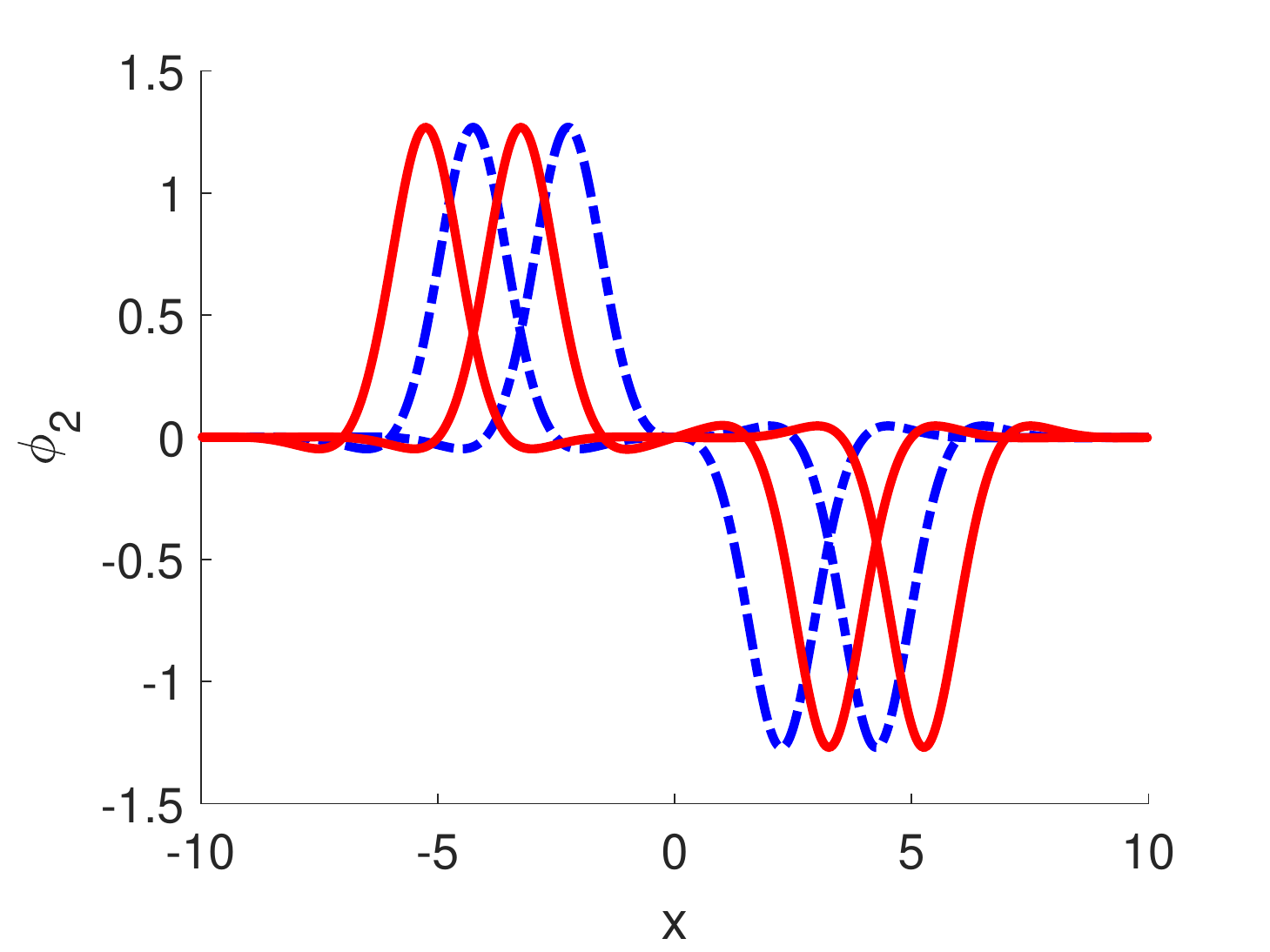}
\end{tabular}
\caption{First four double pulse solutions ($k_1 = 0, 1, 2, 3$) constructed from two pure quartic solitary wave solutions $\phi(x)$ to \cref{standingwavereal}. In-phase double pulses (left panel), opposite phase double pulses (right panel). Dashed blue lines correspond to $k_1$ even, solid red lines correspond to $k_1$ odd. $\beta_2 = 0$, $\beta_4 = -1$, $\omega = 1$, and $\gamma = 1$. }
\label{fig:doublepulses}
\end{figure}

To construct double pulses, we glue together two copies of the primary pulse at the pulse distances predicted by \cref{theorem:multiexist} and solve for the double pulse solution using the same Newton conjugate-gradient method we used above. The first four double pulse solutions are shown in \cref{fig:doublepulses}. 

\begin{figure}[H]
\centering
\begin{tabular}{cc}
\includegraphics[width=8cm]{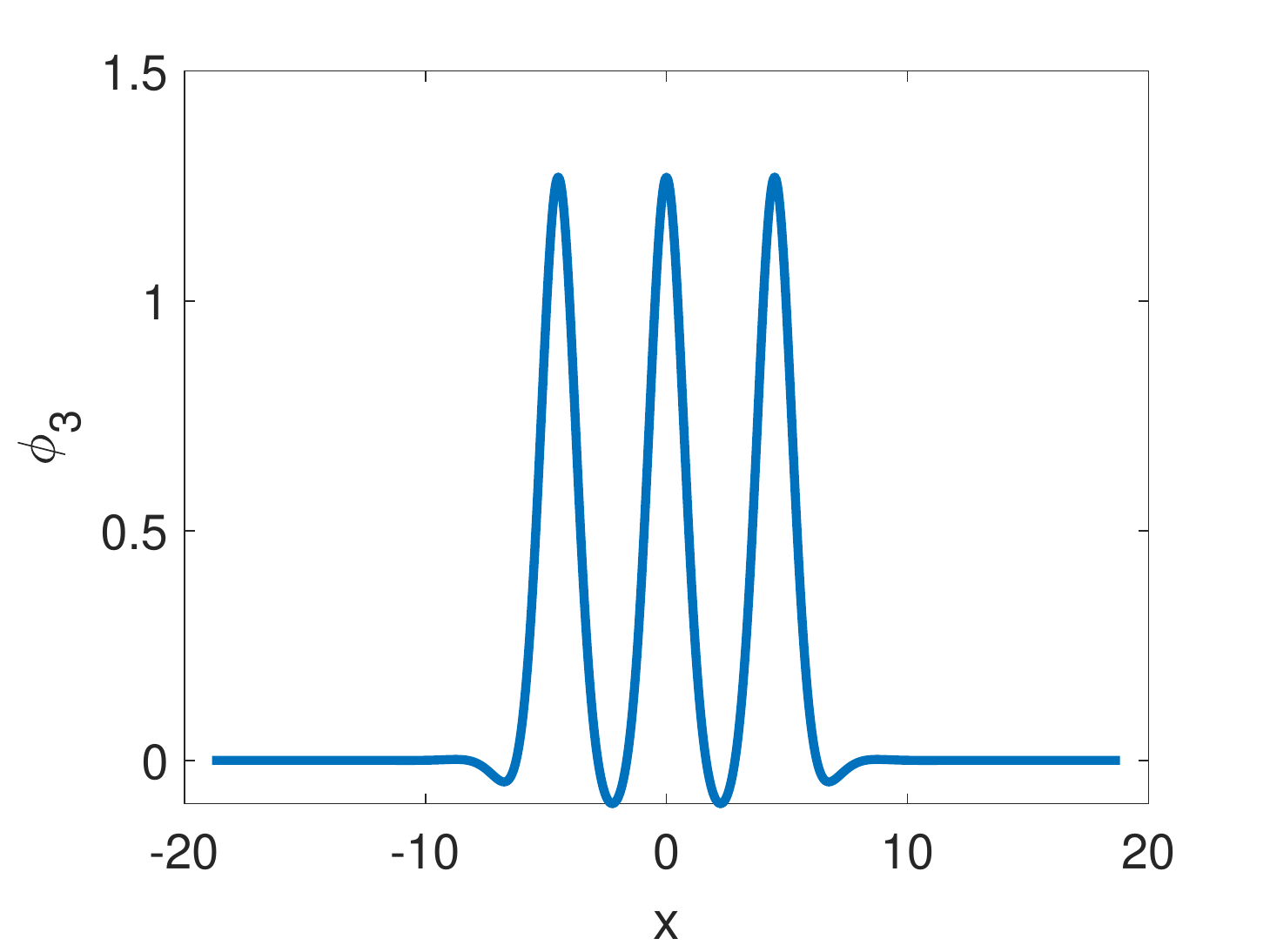} &
\includegraphics[width=8cm]{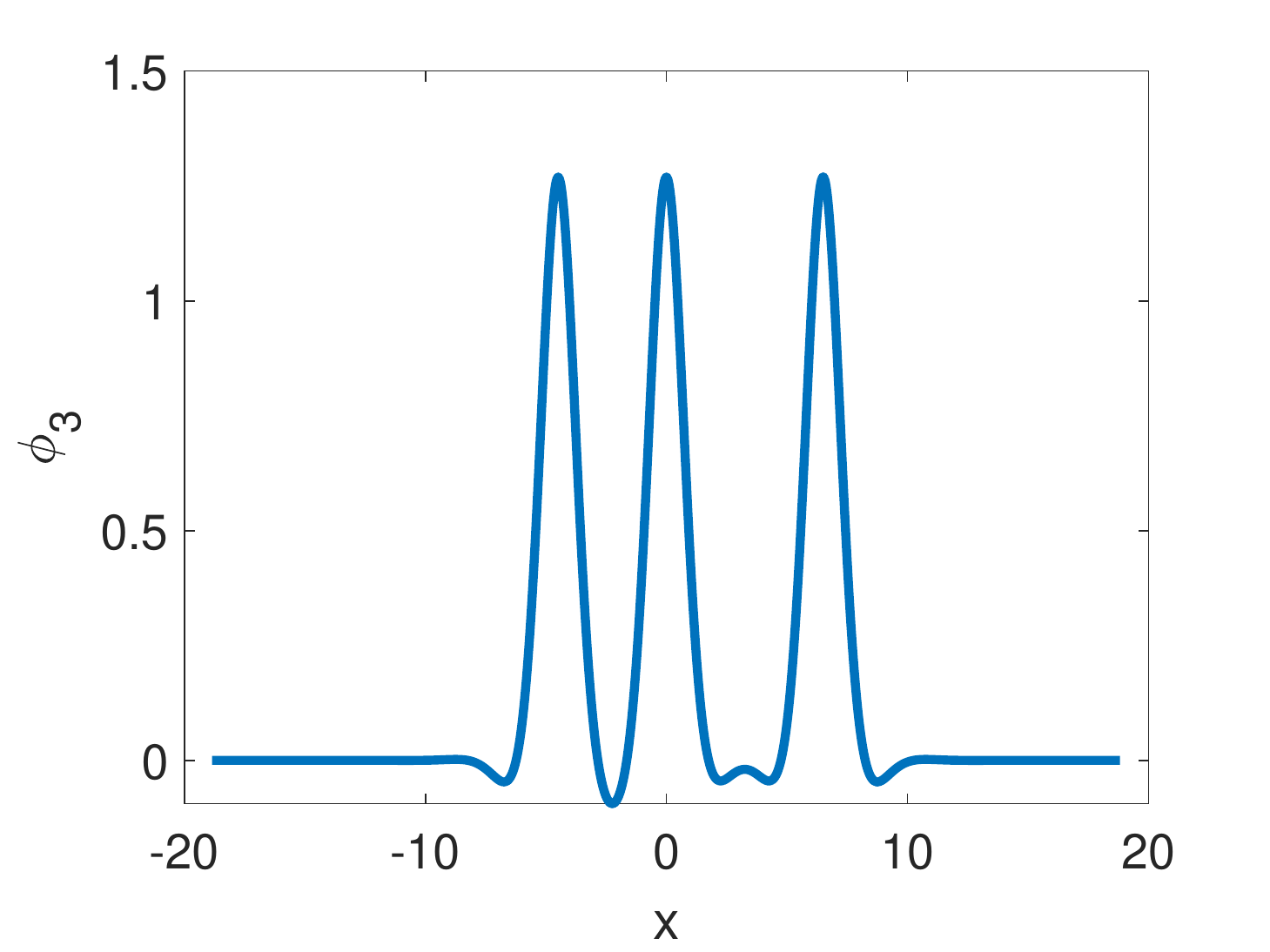}
\end{tabular}
\caption{Triple pulse solutions constructed from three pure quartic solitary wave solutions $\phi(x)$ to \cref{standingwavereal}. Symmetric triple pulse with $(k_1, k_2) = (0,0)$ (left panel), asymmetric triple pulse with $(k_1, k_2) = (0,1)$ (left panel). $\beta_2 = 0$, $\beta_4 = -1$, $\omega = 1$, and $\gamma = 1$.}
\label{fig:triplepulses}
\end{figure}

Arbitrary multi-pulses can similarly be constructed (\cref{fig:triplepulses}). Although the distances between consecutive peaks is constrained by \cref{theorem:multiexist}, these distances do not have to be equal (\cref{fig:triplepulses}, right panel).

\subsection{Spectrum of double pulses}

For both in-phase and out-of-phase double pulses, the spectrum of $J \calL(\phi_2)$, the linearization about the double pulse solution $\phi_2$, contains a pair of purely imaginary interaction eigenvalues and a pair of real interaction eigenvalues, which verifies the result of \cref{corr:2pstab}. The essential spectrum eigenvalues, as expected, are on the imaginary axis and have magnitude $|\lambda| \geq \omega$. There is also a duplication of the internal mode eigenvalues (\cref{fig:doubleinternalmode}); these appear to be purely imaginary, but they do not affect stability since there is always an interaction eigenvalue with positive real part.

\begin{figure}[H]
\centering
\begin{tabular}{cc}
\includegraphics[width=8cm]{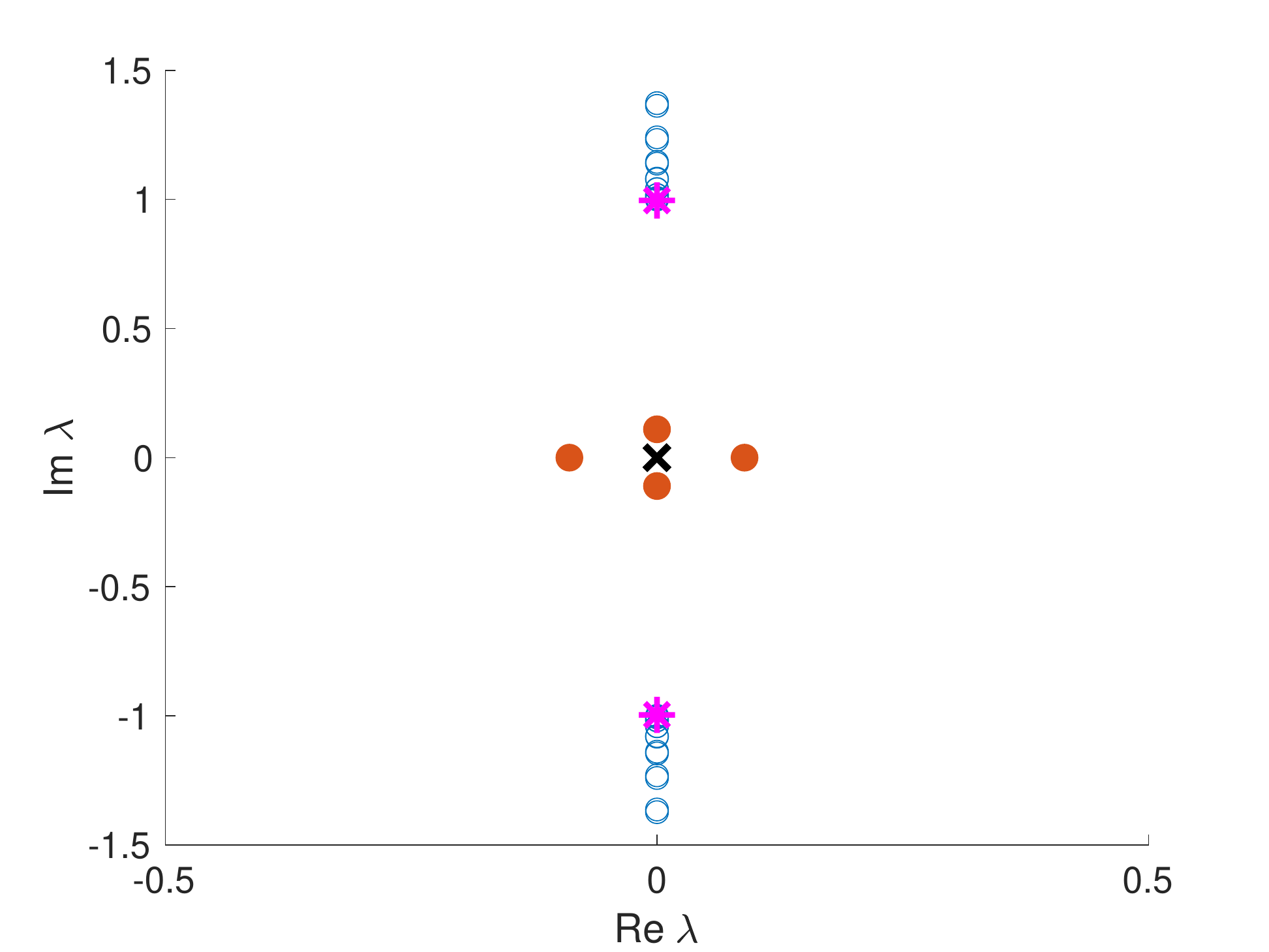} &
\includegraphics[width=8cm]{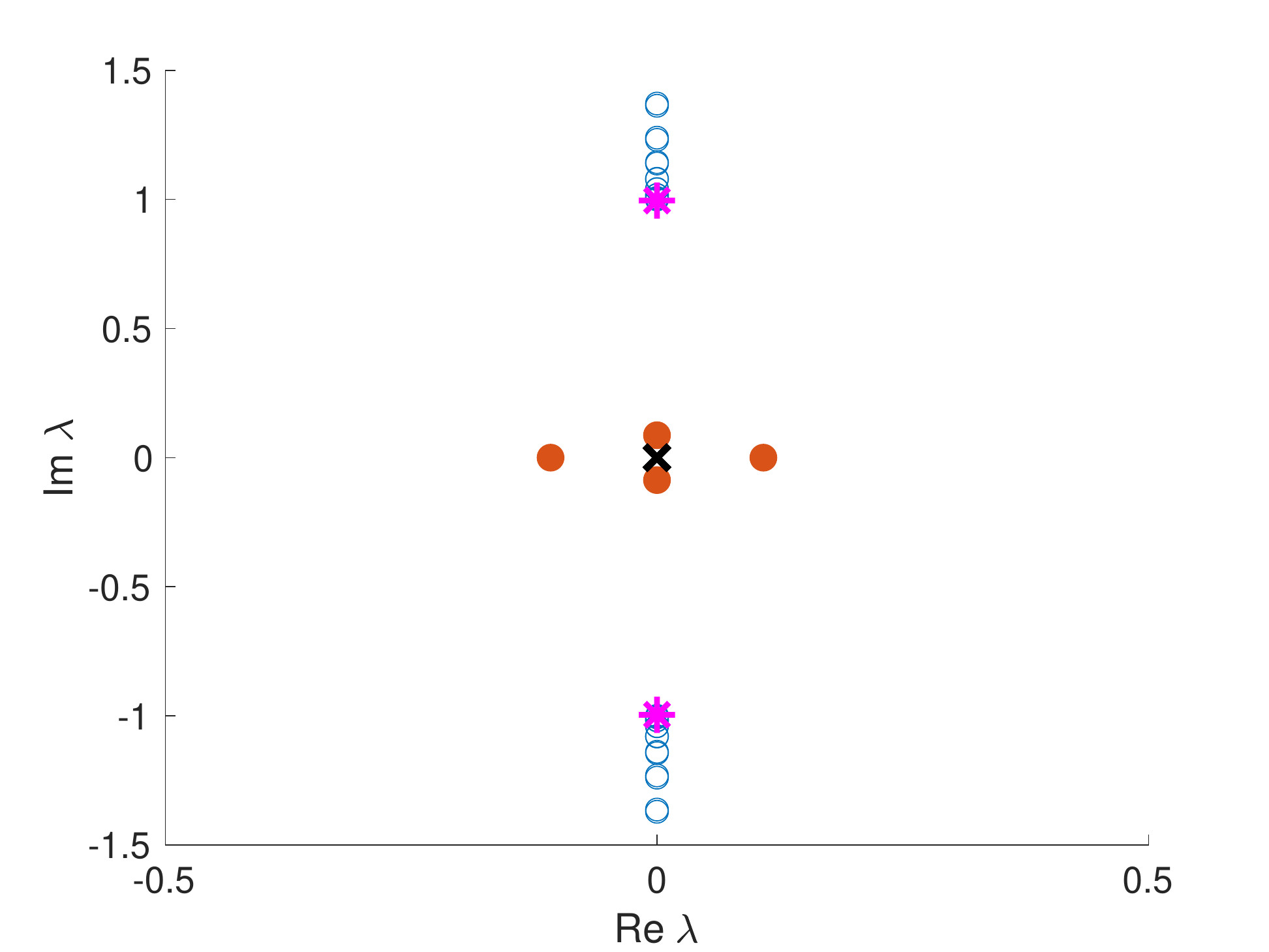}
\end{tabular}
\caption{Eigenvalues for first in-phase double pulse (left panel) and first out-of-phase double pulse (right panel). Spectrum comprises interaction eigenvalues (red dots), essential spectrum (blue open circles), internal mode eigenvalues (magenta stars), and kernel eigenvalue with geometric multiplicity 2 and algebraic multiplicity 4 (black X). See \cref{fig:doublevrvi} for eigenfunctions corresponding to interaction eigenvalues for first in-phase double pulse. $\beta_2 = 0$, $\beta_4 = -1$, $\omega = 1$, and $\gamma = 1$.}
\label{fig:doublespec}
\end{figure} 

\begin{figure}[H]
\centering
\begin{tabular}{c}
\includegraphics[width=8cm]{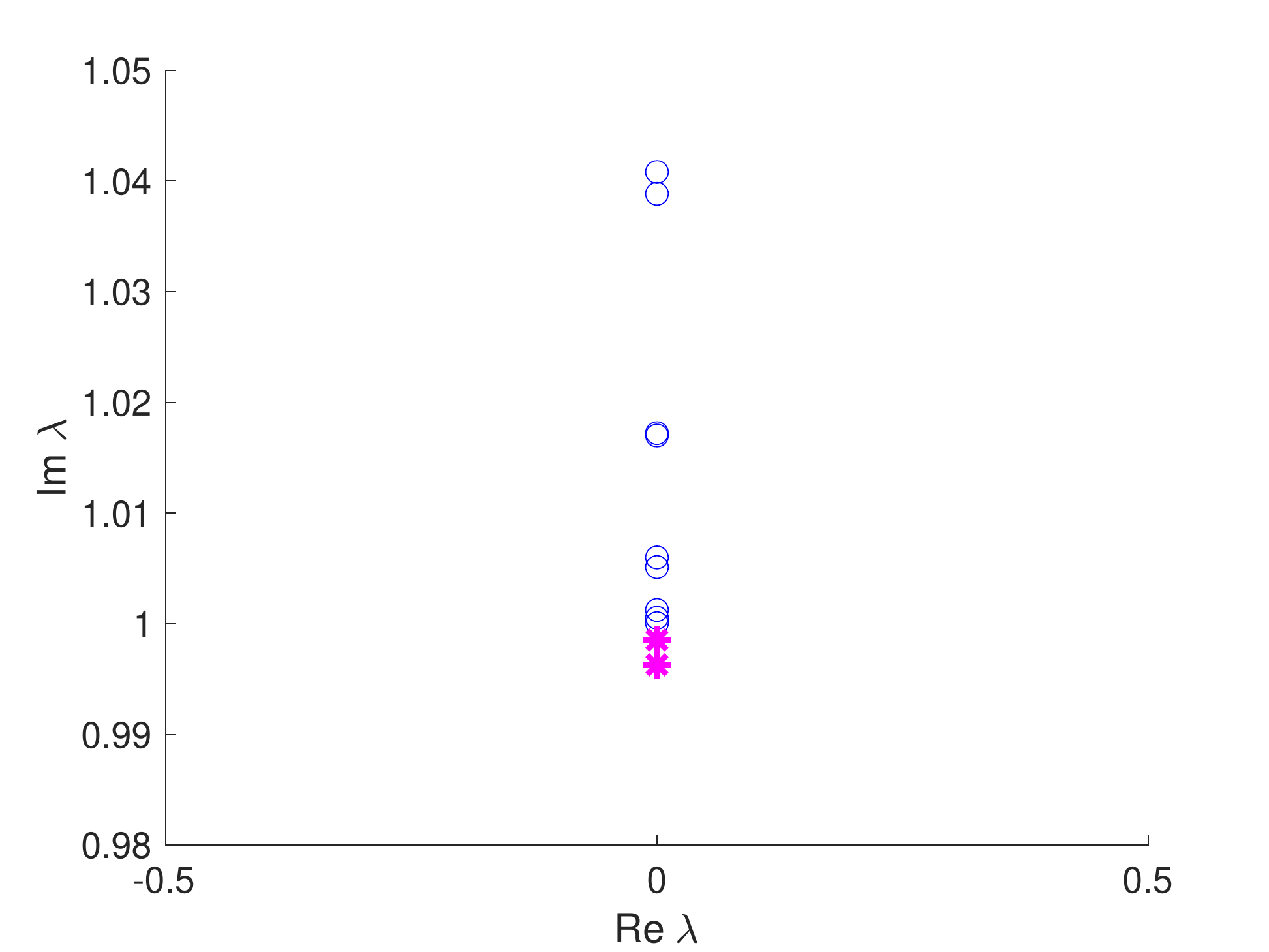}
\end{tabular}
\caption{Close-up of spectrum near $\lambda = i$ showing pair of internal mode eigenvalues (magenta stars) and essential spectrum eigenvalues (blue open circles). Third in-phase double pulse. $\beta_2 = 0$, $\beta_4 = -1$, $\omega = 1$, and $\gamma = 1$.}
\label{fig:doubleinternalmode}
\end{figure}

\begin{figure}[H]
\centering
\begin{tabular}{cc}
\includegraphics[width=8cm]{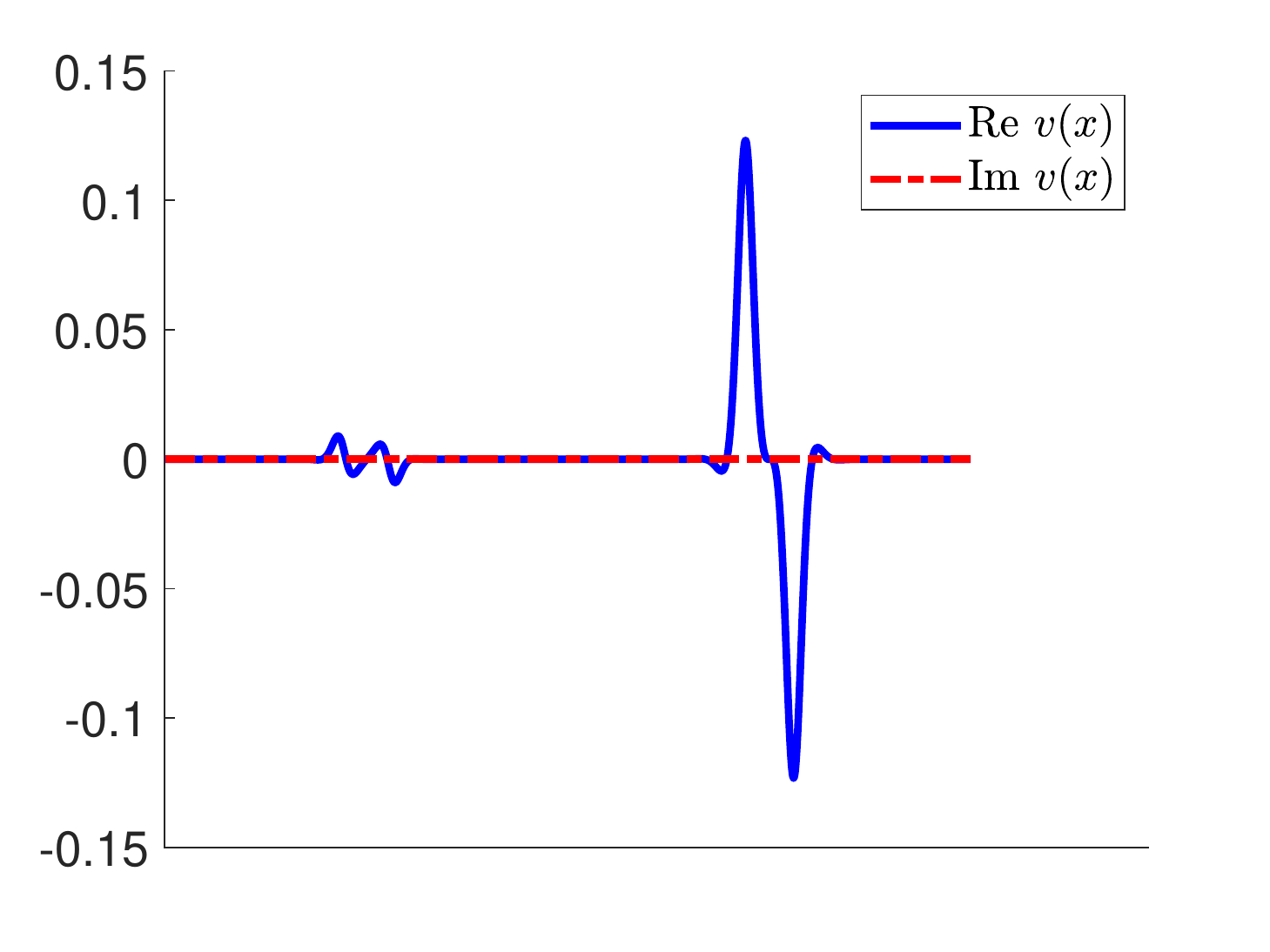} &
\includegraphics[width=8cm]{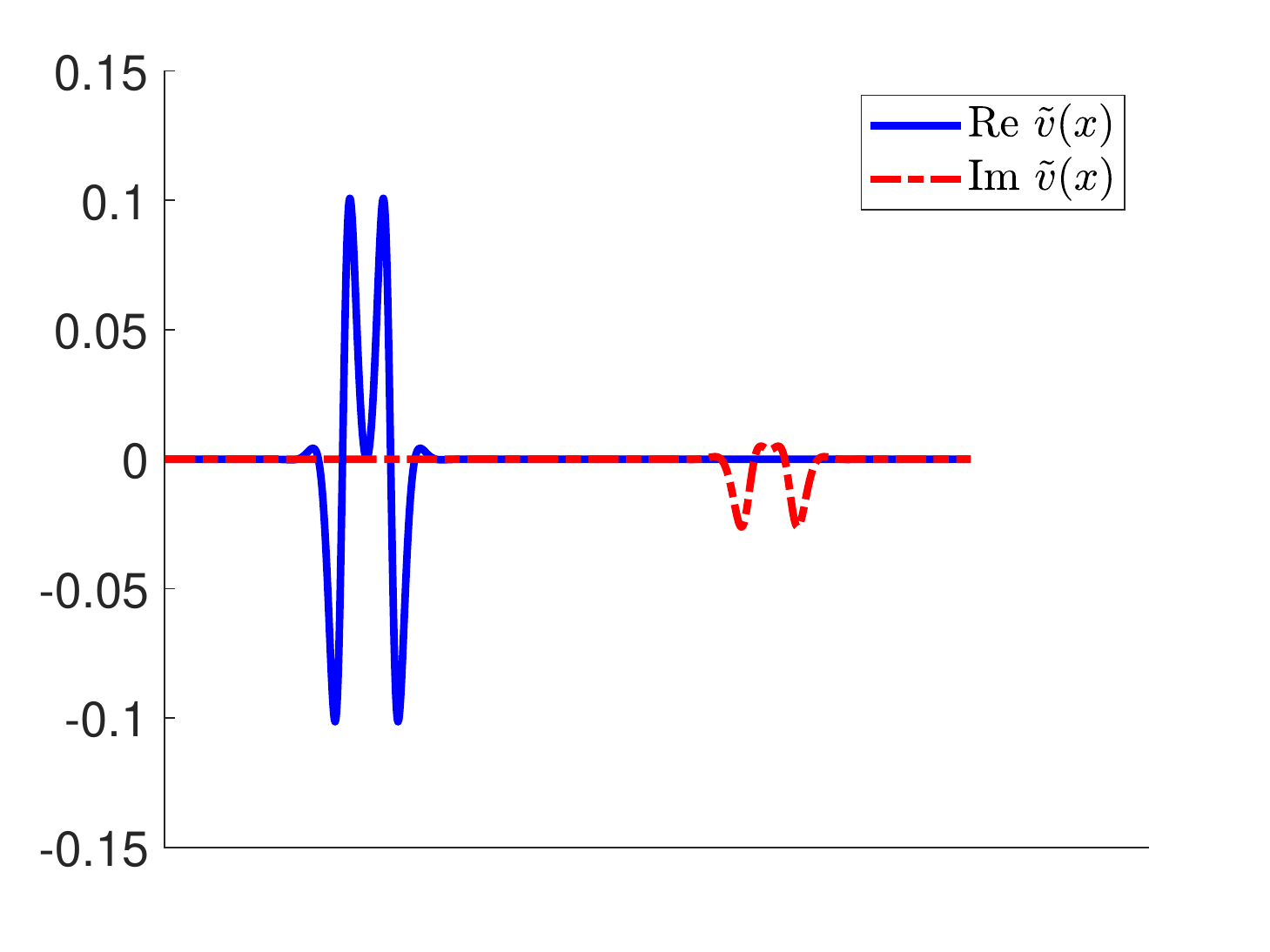}
\end{tabular}
\caption{Eigenfunctions $v(x)$ (left panel) and $\tilde{v}(x)$ (right panel) for first in-phase double pulse $(k_1 = 0)$. $\beta_2 = 0$, $\beta_4 = -1$, $\omega = 1$, and $\gamma = 1$.}
\label{fig:doublevrvi}
\end{figure}

For a double pulse solution $\phi_2$, let $\pm \lambda$ and $\pm \tilde{\lambda}$ be the interaction eigenvalues from \cref{corr:2pstab}, and let $v(x)$ and $\tilde{v}(x)$ be the corresponding eigenfunctions. The eigenfunction $v(x)$ resembles a linear combination of translates of $(0, \phi)^T$, and the eigenfunction $\tilde{v}(x)$ resembles a linear combination of translates of $(\partial_x \phi, 0)^T$ (\cref{fig:doublevrvi}).  (See \cref{sec:proofs} for the construction of these eigenfunctions using Lin's method). For in-phase double pulses, $\lambda$ is real and $\tilde{\lambda}$ is imaginary when $k_0$ is even, and $\lambda$ is imaginary and $\tilde{\lambda}$ is real when $k_0$ is odd. These are reversed for out-of-phase double pulses.

\begin{figure}[H]
\centering
\begin{tabular}{c}
\includegraphics[width=8cm]{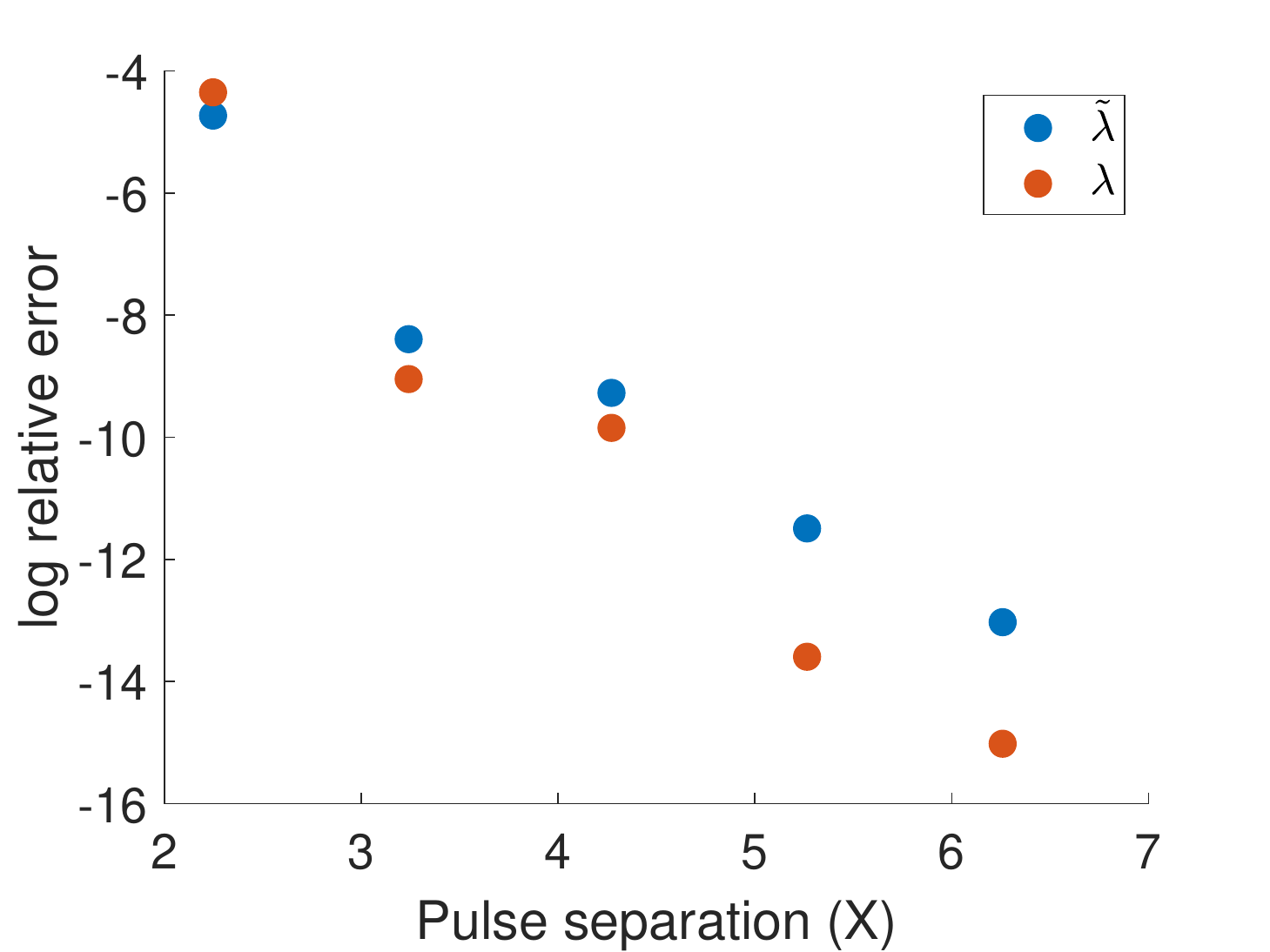}
\end{tabular}
\caption{Log of the relative error for the eigenvalues $\lambda$ and $\tilde{\lambda}$ versus the pulse separation $X$ for the first five in-phase double pulses. $\beta_2 = 0$, $\beta_4 = -1$, $\omega = 1$, and $\gamma = 1$.}
\label{fig:inteigpred}
\end{figure}

Finally, we verify the formulas for the interaction eigenvalues $\lambda$ and $\tilde{\lambda}$ from \cref{corr:2pstab} by plotting the log of the relative error between the leading order term in \cref{inteigpred} and the eigenvalues computed by Matlab versus the pulse separation distance $X$ (\cref{fig:inteigpred}). 

\subsection{Timestepping}

We perform numerical timestepping experiments to characterize the nature of the instability for multi-pulse solutions to \cref{NLS4}. Since the multi-pulses we constructed above are standing waves with frequency $\omega$, we rewrite \cref{NLS4} in a co-rotating frame as 
\begin{equation}\label{NLS4rot}
u_t = i\left( \frac{\beta_4}{24}u_{xxxx} - \frac{\beta_2}{2}u_{xx} - \omega u + \gamma |u|^2 u \right),
\end{equation}
so that the multi-pulses are equilibrium solutions to \cref{NLS4rot}. For the timestepping scheme, we use a split-step Fourier method \cite{Agrawal2013,Bogomolov2006} 
\begin{equation}\label{eq:splitstep}
u(x,t+h) = \calF^{-1} \left\{  e^{i h \left( (\beta_4/24) k^4 - (\beta_2/2)k^2 - \omega \right) } \calF\left( e^{ih |u(x,t)|^2} u(x,t) \right) \right\},
\end{equation}
where $h$ is the time step size, $k$ is the frequency in Fourier space, and the Fourier transform $\calF$ is implemented using the fast Fourier transform.

For double pulse solutions $\phi_2$, based on the eigenfunctions we computed in the previous section, we expect to see perturbations evolve either in the distance between the two peaks, corresponding to the eigenfunction $\tilde{v}(x)$, or in the phase difference between the two peaks, corresponding to the eigenfunction $v(x)$. (See \cite[Figure 9]{Pelinovsky2007} for timestepping results for double pulses in the 5th order KdV equation; in that case, there is a single continuous symmetry in the underlying system, and perturbations evolve in the distance between the two peaks).

\begin{figure}
\centering
\begin{tabular}{cc}
\includegraphics[width=8cm]{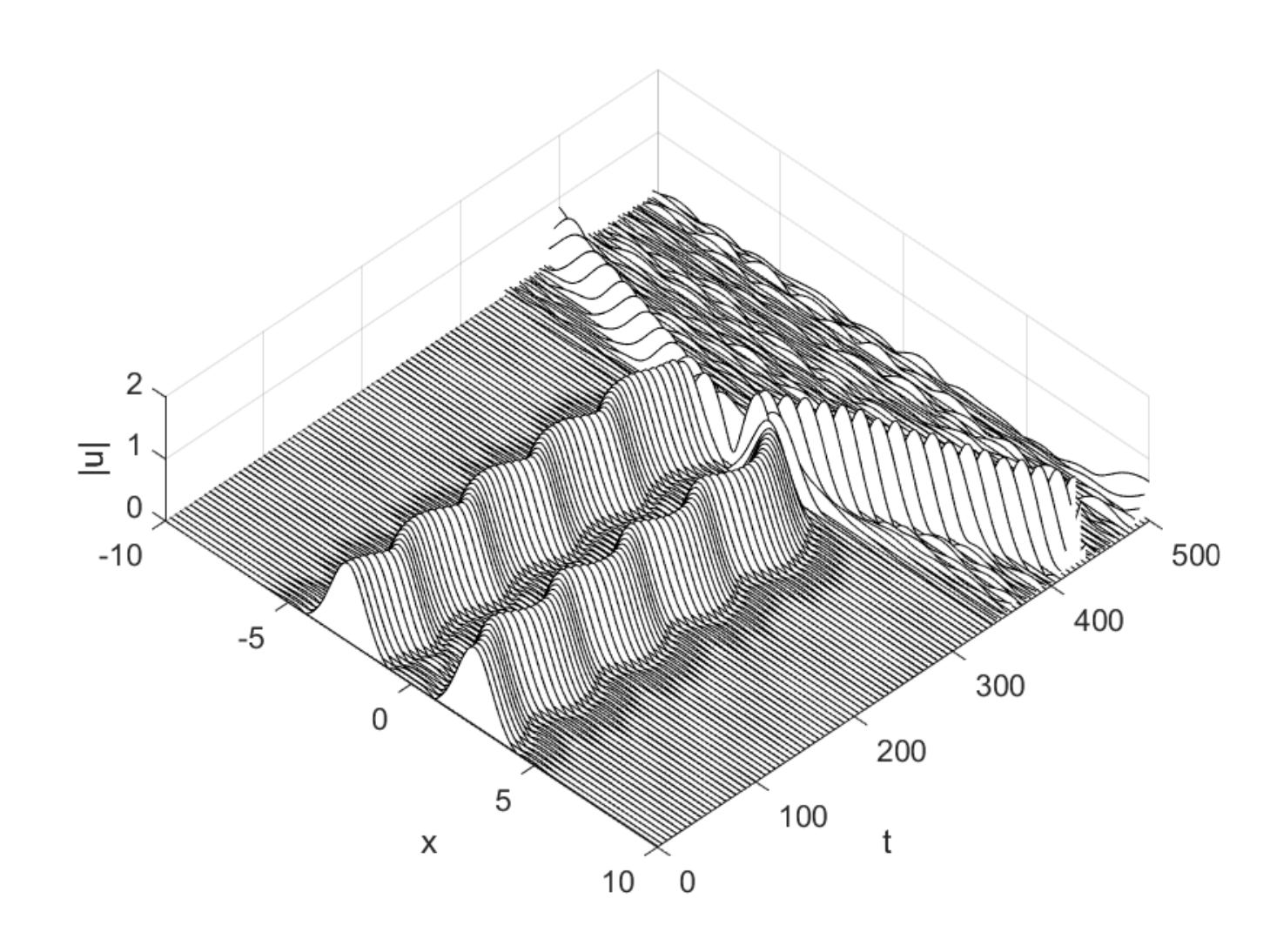} &
\includegraphics[width=8cm]{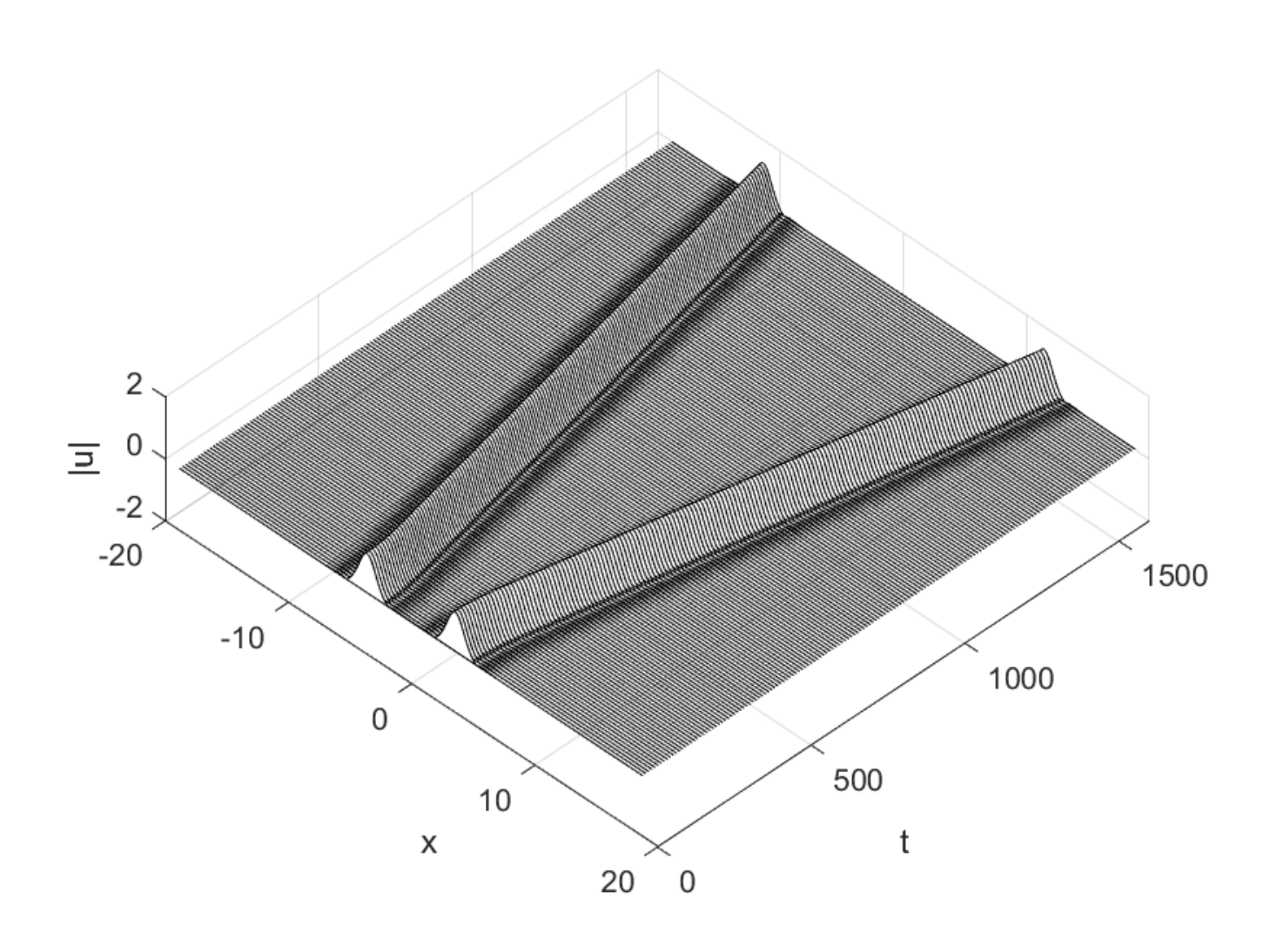} \\
\includegraphics[width=8cm]{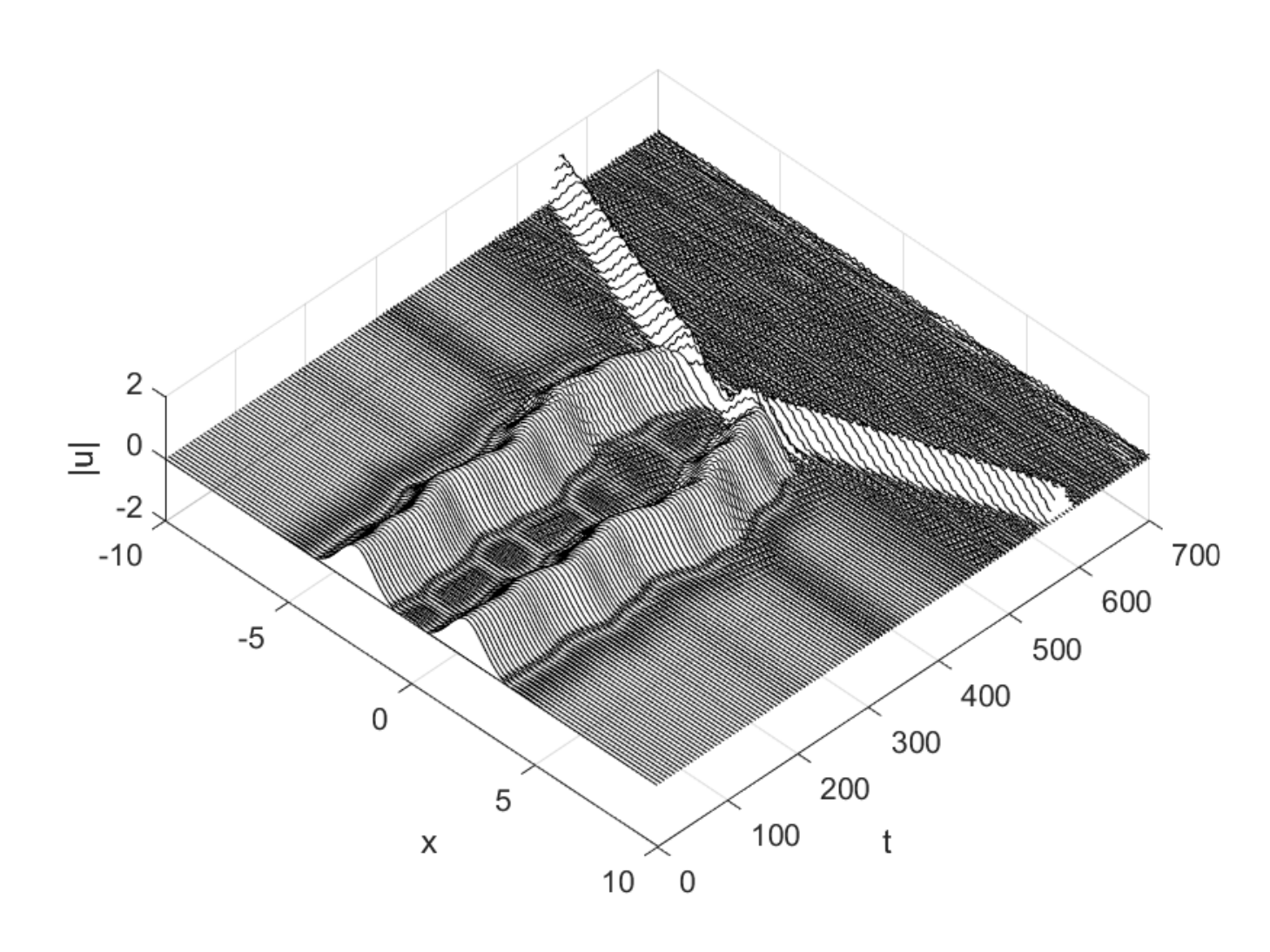} &
\includegraphics[width=8cm]{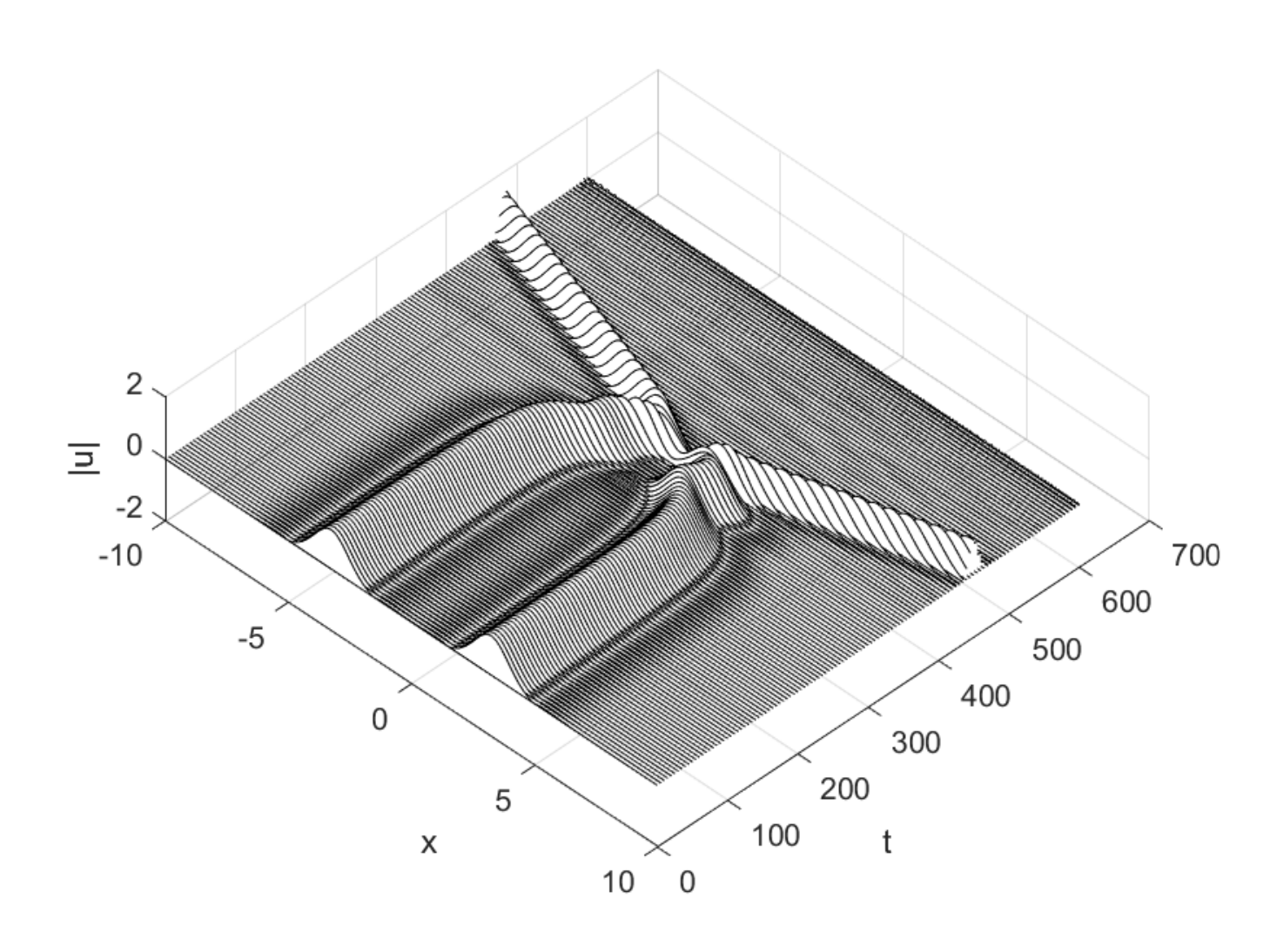} \\
\includegraphics[width=6cm]{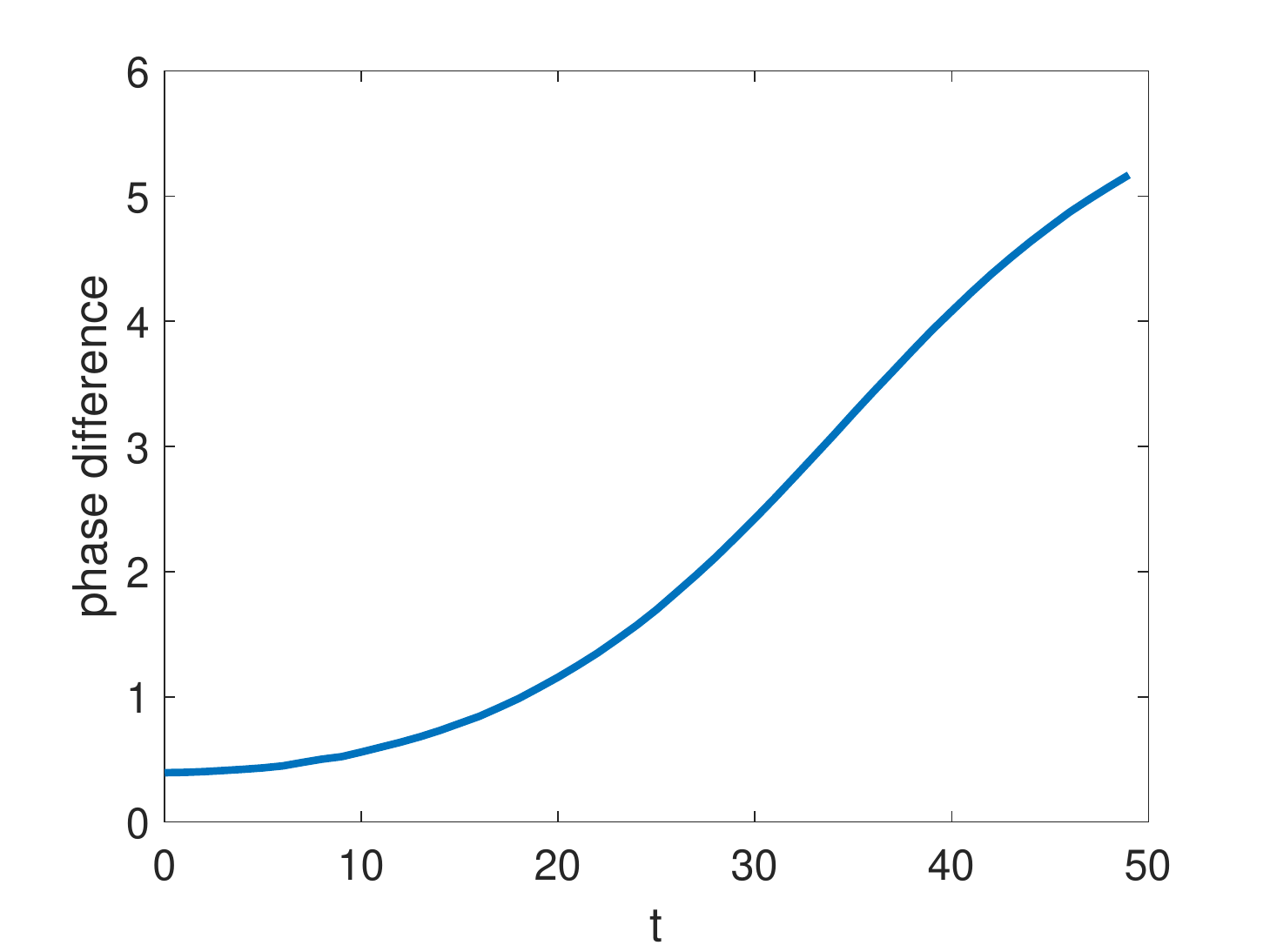} &
\includegraphics[width=6cm]{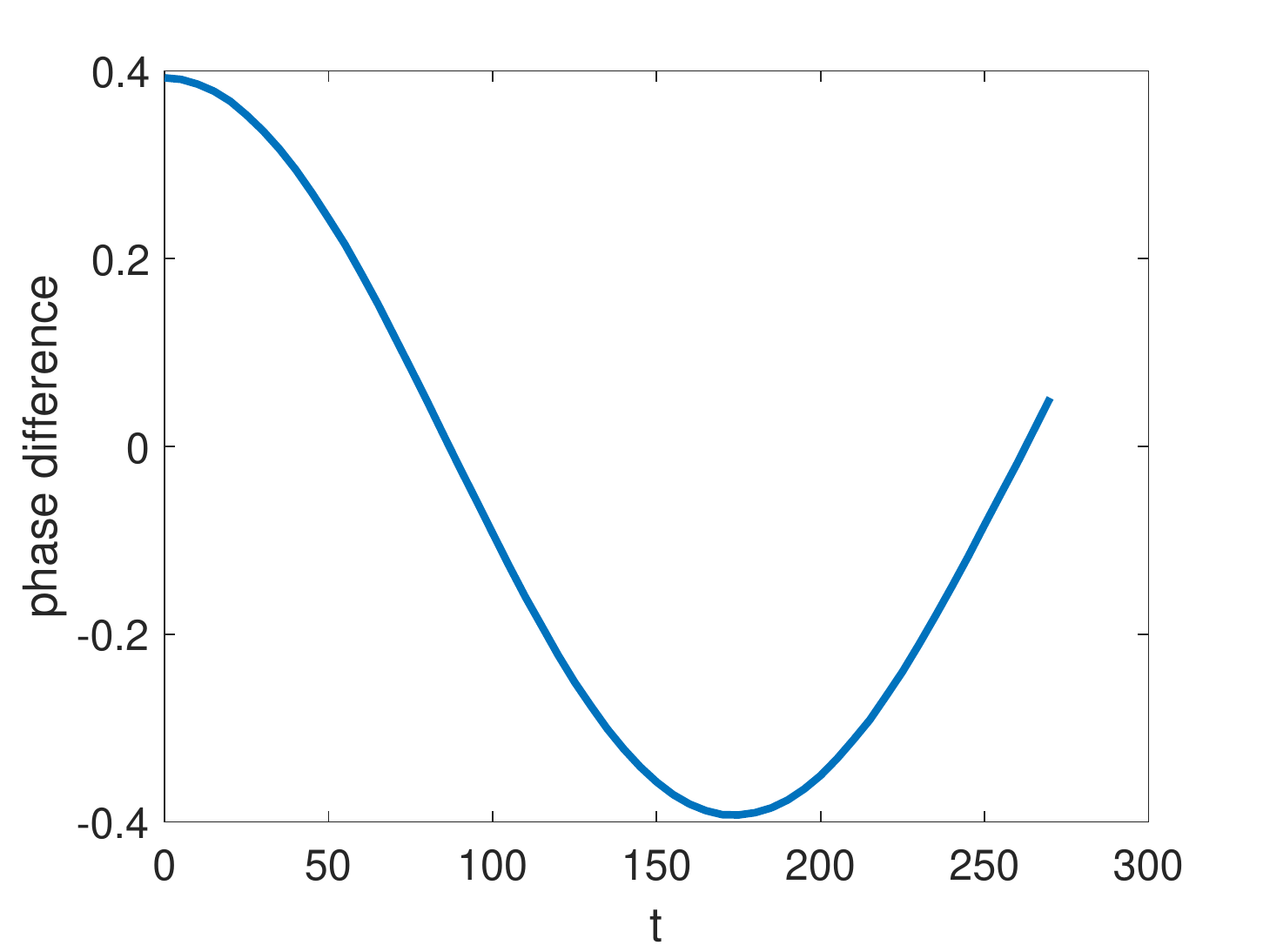}
\end{tabular}
\caption{Timestepping results for perturbations of double pulses. Left column is first in-phase double pulse ($k_1 = 0$, interaction eigenvalues $\lambda = \pm 0.0891$ and $\tilde{\lambda} = \pm 0.1098 i$) and right column is second in-phase double pulse ($k_1 = 1$, interaction eigenvalues $\lambda = \pm 0.0183i$ and $\tilde{\lambda} = \pm 0.0227 i$). Top row is evolution of norm of solution perturbed by increasing inter-peak distance by $0.3680$ (10 grid points). Middle row is evolution of norm of solution perturbed by rotating peaks in opposite directions to attain a phase difference of $\pi/8$; the evolution of this phase difference is shown in the bottom row. Time evolution by split-step Fourier method \cref{eq:splitstep} with time step $h = 0.001$. $\beta_2 = 0$, $\beta_4 = -1$, $\omega = 1$, and $\gamma = 1$.}
\label{fig:timestep0pp}
\end{figure}

\cref{fig:timestep0pp} shows timestepping results for the first and second in-phase double pulses. For $k_0 = 0$ (\cref{fig:timestep0pp}, left column), $\lambda$ is real and $\tilde{\lambda}$ is imaginary. When the two peaks are pulled apart (\cref{fig:timestep0pp}, top left), they oscillate briefly about the equilibrium position before the structure loses stability. The angular frequency of these oscillations (approximately $0.1067$) is within 2\% of the imaginary part of the small, purely imaginary eigenvalue ($0.1098i$), which indicates that these oscillations are the result of this interaction eigenvalue rather than the internal mode eigenvalue of the base pulse. When the two peaks are rotated in opposite directions (\cref{fig:timestep0pp}, middle and bottom left), the phase difference between the peaks continues to grow. 

For $k_0 = 1$ (\cref{fig:timestep0pp}, right column), $\lambda$ is imaginary and $\tilde{\lambda}$ is real. When the two peaks are pulled apart (\cref{fig:timestep0pp}, top right), they repel each other and travel in opposite directions with equal speeds. This behavior resembles that of a pair of solitons. (See the top left and bottom panels of \cite[Figure 9]{Pelinovsky2007} for similar behavior in the 5th order KdV equation). When the two peaks are rotated in opposite directions (\cref{fig:timestep0pp}, middle and bottom right), the phase difference between peaks briefly oscillates about 0 before the structure loses stability. The angular frequency of these phase oscillations (approximately $0.0180$) is also within 2\% of the imaginary part of the small, purely imaginary eigenvalue ($0.0183i$), which indicates that these oscillations are the result of this interaction eigenvalue. For in-phase double pulses with $k$ even, the timestepping results are similar to those in the left column of \cref{fig:timestep0pp}, and for in-phase double pulses with $k$ odd, the timestepping results are similar to those in the right column of \cref{fig:timestep0pp}. This is reversed for out-of-phase double pulses.

\section{Conclusions and future directions}

In this paper, we studied single and multi-pulse solitary wave solutions to a general nonlinear Schr{\"o}dinger equation with both second and fourth-order dispersion terms. We first gave criteria for the existence of a primary soliton solution in terms of the parameters of the system, and provided numerical verification for hypotheses leading to orbital stability of the primary pulse. For the classical NLS equation, \cref{hyp:Lminusspec} follows from Sturm-Liouville theory. It may be possible to use a tool such as the Maslov index to prove this result for the fourth-order equation (see, for example, \cite{Chardard2009,Chardard2011,Jones1988}). The Maslov index could also be used as an alternative technique for characterizing the spectra of multi-pulses \cite{Chardard2009a}. Our theoretical results compare well with experimental pulses reported in \cite{BlancoPQS} in two key aspects. From \cite[Figure 2b]{BlancoPQS} (for power 0.7W), the flat phase across the pulse is consistent with the theoretical ansatz leading to equation \cref{standingwavereal}, and oscillations observed in the experimental frequency domain are consistent with the theoretical proof of oscillatory tails. To extend the model to more accurately capture experimental results would require adding linear and nonlinear loss terms. More recent experimental work in the setting of laser optics \cite{Blanco_laser} overcomes these losses. Here, the authors demonstrate that the energy of these pulses is proportional to the third power of the inverse pulse duration, which is consistent to the scaling we use leading to equation \cref{eq:NLS4rescaled}.   

We then constructed $n$-pulse solutions by splicing together multiple copies of the primary pulse, and reduced the problem of finding the small eigenvalues resulting from interaction between neighboring pulses to that of computing the determinant of a $2n\times2n$ block matrix. Under the same assumptions which lead to orbital stability of the primary pulse, we showed that all multi-pulse solutions are unstable. To our knowledge, there are no experimental results on $n$-solitons. It would be interesting to see if future work on lasers \cite{Blanco_laser} could produce $n$-solitons with cavity round-trip separation that would allow our instability predictions to be verified. It would also be interesting to see experimental results on the propagation of pulse trains where the distance between subsequent pulses is small. In future research we could investigate solitons and multi-pulses in higher order NLS equations, as discussed in \cite{Runge2020}. We expect that these results would hold for these higher order variants, and that all multi-pulse solutions would similarly be unstable. We could also study generalizations to other nonlinearities. Finally, this equation represents an idealization of the experimental situation since energy is always conserved. A more realistic model might incorporate gains and losses of energy in the laser cavity, and it is possible that stable multi-pulses could exist in such a scenario.

\section{Proof of stability results}\label{sec:proofs}

\subsection{Proof of Theorem \ref{th:blockmatrix}}\label{sec:blockmatrixproof}

The proof is adapted from \cite[Section 3.4]{Manukian} and the proof of \cite[Theorem 2]{Sandstede1998}, and uses an implementation of the Lyapunov-Schmidt reduction known as Lin's method. It follows from \cref{Lphikernel} that
\begin{equation}\label{Kkernel}
\begin{aligned}
[Y(x)]' &= K(\phi_n)Y(x), \quad [Z(x)]' = K(\phi_n)Z(x) + B_1 Y(x) \\
[\tilde{Y}(x)]' &= K(\phi_n)\tilde{Y}(x), \quad [\tilde{Z}(x)]' = K(\phi_n)\tilde{Z}(x) + B_1 \tilde{Y}(x),
\end{aligned}
\end{equation}
where
\begin{equation}
\begin{aligned}
Y(x) &= ( 0, U_n(x) )^T, \quad
Z(x) = ( \partial_\omega U_n(x), 0 )^T \\
\tilde{Y}(x) &= ( \partial_x U_n(x), 0)^T, \quad
\tilde{Z}(x) = ( 0, Z_n(x) )^T,
\end{aligned}
\end{equation}
$Z_n(x) = (z_n(x), \partial_x z_n(x), \partial_x^2 z_n(x), \frac{\beta_4}{24} \partial_x^3 z_n(x))$, and the first component $z_n(x)$ solves $\calL^-(\phi_n)z_n = \phi_n'$. The analysis is identical to that of \cite{Manukian}, except the piecewise ansatz for the eigenfunction also involves $Z(x)$ and $\tilde{Z}(x)$ as in \cite{Parker2020}. Writing the functions \cref{Kkernel} in piecewise form as with \cref{Unpiecewise}, we take the ansatz
\begin{align}\label{Vansatz}
V_i^\pm(x)' &= d_i(Y_i^\pm(x) + \lambda Z_i^\pm(x)) + \tilde{d}_i(\tilde{Y}_i^\pm(x) + \lambda \tilde{Z}_i^\pm(x)) + W_i^\pm && i = 1, \dots, n,
\end{align}
where $V_i^- \in C^0( [-X_{i-1}, 0], \C^8 )$ and $V_i^+ \in C^0( [0, X_i], \C^8 )$. Substituting \cref{Vansatz} into \cref{multieig} and simplifying using \cref{Kkernel}, the remainder functions $W_i^\pm(x)$ solve the equation
\begin{align}\label{Wsolves}
W_i^\pm(x)' &= K(\phi_n)W_i^\pm(x) + \lambda^2 d_i B Z_i^\pm(x) + \lambda^2 \tilde{d}_i B \tilde{Z}_i^\pm(x) && i = 1, \dots, n.
\end{align}
Following \cite{Manukian,Sandstede1998}, we obtain a unique piecewise solution $W_i^\pm(x)$ which generically has $n$ jumps at $x = 0$ in the direction of $Q^*(0) \oplus \tilde{Q}^*(0)$. Using the definitions of $Q^*(x)$ and $\tilde{Q}^*(x)$ together with \cref{Unestimates} and \cite[(3.19)]{Manukian}, these jumps are given by
\begin{equation}\label{jumpcond1}
\begin{aligned}
\xi_i &= \theta_{i+1} \langle \Psi(X_i), U(-X_i) \rangle (d_{i+1} - d_i) 
+ \theta_{i-1} \langle \Psi(-X_{i-1}), U(X_{i-1}) \rangle (d_i - d_{i-1} )  \\
&\qquad + \lambda^2 \theta_i d_i \int_{-\infty}^\infty \langle \Psi(y), B \partial_\omega U(y) \rangle dy 
+ \mathcal{O}((|\lambda| + e^{-\alpha X_{\min}})^3) \\
\tilde{\xi}_i &= \theta_{i+1} \langle \Psi'(X_i), U'(-X_i) \rangle (\tilde{d}_{i+1} - \tilde{d}_i) 
+ \theta_{i-1} \langle \Psi'(-X_{i-1}), U'(X_{i-1}) \rangle (\tilde{d}_i - \tilde{d}_{i-1}) \\
&\qquad- \lambda^2 \theta_i \tilde{d}_i \int_{-\infty}^\infty \langle \Psi(y), B Z(y) \rangle dy 
+ \mathcal{O}((|\lambda| + e^{-\alpha X_{\min}})^3),
\end{aligned}
\end{equation}
where $Z(x) = (z(x), \partial_x z(x), \partial_x^2 z(x), \frac{\beta_4}{24} \partial_x^3 z(x))$. By symmetry, 
\begin{equation}\label{Rrelation}
\Psi(-x) = -R \Psi(x), \quad U(-x) = R U(x),
\end{equation}
where $R$ is the standard reversor operator 
\[
R(u_1, u_2, u_3, u_4) = (u_1, -u_2, u_3, -u_4),
\] 
thus 
\begin{equation}\label{PsiR}
\begin{aligned}
\langle \Psi(-X_{i-1}), U(X_{i-1}) \rangle &= -\langle \Psi(X_{i-1}), U(-X_{i-1}) \rangle \\
\langle \Psi'(-X_{i-1}), U'(X_{i-1}) \rangle &= -\langle \Psi'(X_{i-1}), U'(-X_{i-1}) \rangle.
\end{aligned} 
\end{equation}
Finally, we relate $\langle \Psi(X_i), U(-X_i) \rangle$ and $\langle \Psi'(X_i), U'(-X_i) \rangle$. Since $DF(\phi) = K^+(\phi)$, $\Psi(x)$ is the unique bounded solution to the adjoint equation $W'(x) = -DF(\phi)^* W(x)$. Thus by \cite[Lemma 6.1]{Sandstede1998}, with $\Psi'(x)$ in place of $\Psi(x)$, $p$ in place of $\phi$, and no parameter $\mu$,
\begin{align}
\langle \Psi'(x), U(-x) &= \langle \Psi'(-x), U(x) \rangle = s e^{-2 a x} \sin(2 b x + p) + \mathcal{O}(e^{-(2 \alpha + \gamma)x}) \label{IPdPsiU} \\
\langle \Psi'(x), U'(-x) \rangle &= 
\langle \Psi'(x), U'(-x) \rangle = -s e^{-2 a x} \left( b \cos(2 b x + p) - a \sin(2 b x + p) \right) + \mathcal{O}(e^{-(2 \alpha + \gamma)x}), \label{IPdPsidU}
\end{align}
where $s > 0$ and $\gamma > 0$. Differentiating $\langle \Psi(-x), U(x) \rangle$ with respect to $x$, since the operator $\partial_x$ is skew symmetric, 
\begin{align*}
\frac{d}{dx} \langle \Psi(x), U(-x) \rangle = 2 \langle \Psi'(x), U(-x) \rangle,
\end{align*}
thus we can integrate \cref{IPdPsiU} by parts to get 
\begin{align*}
\langle \Psi(x), U(-x) \rangle = -\frac{1}{a^2 + b^2} s e^{-2 a x} \left( b \cos(2 b x + p) + a \sin(2 b x + p) \right) + \mathcal{O}(e^{-(2 \alpha + \gamma)x}).
\end{align*}
In the proof of \cite[Theorem 3]{Sandstede1998}, the distances $X_i$ are chosen to solve $s e^{-2 a X_i} \sin(2 b X_i + p) = \mathcal{O}(e^{-(2 \alpha + \gamma)X_i})$, thus for $x = X_i$ we have
\begin{align}\label{IPrelation1}
\langle \Psi'(X_i), U'(-X_i) \rangle = (a^2 + b^2)
\langle \Psi(X_i), U(-X_i) \rangle + \mathcal{O}(e^{-(2 \alpha + \gamma)X_i}).
\end{align}
Using \cref{IPrelation1} and \cref{PsiR}, multiplying by $\theta_i$, and using the definition of $B$, equations \cref{jumpcond1} simplify to the jump conditions
\begin{align*}
\xi_i &= \theta_i \theta_{i+1} \langle \Psi(X_i), U(-X_i) \rangle (d_{i+1} - d_i) 
- \theta_{i-1} \theta_i  \langle \Psi(X_{i-1}), U(-X_{i-1}) \rangle (d_i - d_{i-1}) \\
&\qquad + \lambda^2 d_i \int_{-\infty}^\infty \phi(y) \partial_\omega \phi(y) dy 
+ \mathcal{O}( |\lambda|(|\lambda| + e^{-\alpha X_{\min}})^2 + e^{-(2 \alpha + \gamma)X_{\min} }) )  \\
\tilde{\xi}_i &= (a^2 + b^2) \theta_i \theta_{i+1} \langle \Psi(X_i), U(-X_i) \rangle (\tilde{d}_{i+1} - \tilde{d}_i)
- (a^2 + b^2) \theta_{i-1} \theta_i \langle \Psi(X_{i-1}), U(-X_{i-1}) \rangle (\tilde{d}_i - \tilde{d}_{i-1}) \\
&\qquad- \lambda^2 \tilde{d}_i \int_{-\infty}^\infty \partial_y \phi(y) z(y) dy
+ \mathcal{O}( |\lambda|(|\lambda| + e^{-\alpha X_{\min}})^2 + e^{-(2 \alpha + \gamma)X_{\min} }) ) ,
\end{align*}
which we write in matrix form as in the statement of the theorem.

\subsection{Proof of Corollary \ref{corr:multiunstable} and Corollary \ref{corr:2pstab}}

For \cref{corr:multiunstable}, let $\{ \mu_1,\dots,\mu_{n-1}, 0\}$ be the eigenvalues of $A$, which are real and distinct as in the proof of \cite[Theorem 5]{Parker2020}. Following the steps in that proof and using the rescaling in \cite[Theorem 3]{Sandstede1998}, there are $2(n-1)$ pairs of interaction eigenvalues, given by \cref{inteigs}, which are either real or purely imaginary by Hamiltonian symmetry. There is also an eigenvalue with algebraic multiplicity 4 at the origin. Since $M > 0$ and $\tilde{M} > 0$ by \cref{hyp:Mcond}, one of each pair $\lambda_i, \tilde{\lambda}_i$ is real and the other is purely imaginary. \cref{corr:2pstab} is the specific case $n = 2$, where the nonzero eigenvalue of $A$ can be computed directly.

\subsubsection*{Acknowledgments}

This material is based upon work supported by the U.S. National Science Foundation under the RTG grant DMS-1840260 (R.P. and A.A.).

\bibliography{nls4.bib}

\end{document}